\documentclass[pra,superscriptaddress,twocolumn,floatfig,showpacs,floatfix]{revtex4-1}
\usepackage{graphicx,color}
\usepackage{amsmath,amssymb,bm}
\usepackage{braket}		
\usepackage{multirow}
\DeclareMathOperator{\sgn}{sgn}

\newcommand{\la}{\Lambda}
\newcommand{\tn}{\textnormal}

\definecolor{darkred}{rgb}{0.90,0,0}
\definecolor{darkgreen}{rgb}{0,0.60,.2}
\definecolor{darkblue}{rgb}{0,0,1}
\definecolor{grey}{cmyk}{0,0,0,0.25}
\definecolor{orange}{cmyk}{0,0.6,1,0}

\begin{document}
\title{Many-body localization of spinless fermions with attractive interactions in one dimension}

\author{Sheng-Hsuan Lin}
\affiliation{Department of Informatics, Technische Universit\"at M\"unchen, 85748 Garching, Germany}
\affiliation{Department of Physics and Arnold Sommerfeld Center for Theoretical Physics,
Ludwig-Maximilians-Universit\"at M\"unchen, D-80333 M\"unchen, Germany}
\author{B. Sbierski}
\affiliation{Dahlem Center for Complex Quantum Systems and Fachbereich Physik, Freie Universit\"at Berlin, 14195 Berlin, Germany}
\author{F. Dorfner}
\affiliation{Department of Physics and Arnold Sommerfeld Center for Theoretical Physics,
Ludwig-Maximilians-Universit\"at M\"unchen, D-80333 M\"unchen, Germany}
\author{C. Karrasch}
\affiliation{Dahlem Center for Complex Quantum Systems and Fachbereich Physik, Freie Universit\"at Berlin, 14195 Berlin, Germany}
\author{F. Heidrich-Meisner}
\affiliation{Department of Physics and Arnold Sommerfeld Center for Theoretical Physics,
Ludwig-Maximilians-Universit\"at M\"unchen, D-80333 M\"unchen, Germany}

\begin{abstract}
We study the finite-energy density phase diagram of spinless fermions with attractive interactions in one dimension in the presence
of uncorrelated diagonal disorder. Unlike the case of repulsive interactions, a delocalized Luttinger-liquid phase persists at weak disorder
in the ground state, which is a well-known result. We revisit the ground-state phase diagram and show that the recently introduced occupation-spectrum discontinuity 
computed from the eigenspectrum of one-particle density matrices is noticeably smaller in the Luttinger liquid compared to the
localized regions. Moreover, we use the functional renormalization scheme to study the finite-size dependence
of the conductance, which resolves the existence of the Luttinger liquid as well and is computationally cheap.
Our main results concern the finite-energy density case. Using exact diagonalization and by computing various established measures of the 
many-body localization-delocalization transition, we argue that the zero-temperature Luttinger liquid smoothly evolves into a finite-energy density
ergodic phase without any intermediate phase transition. 
\end{abstract}

\maketitle



\section{Introduction}
\label{sec:introduction}
A much studied model for the interplay between disorder and interactions at finite energy densities are spinless fermions
in a one-dimensional lattice with uncorrelated diagonal disorder and nearest-neighbor repulsive interactions.
The Hamiltonian, defined on a one-dimensional (1D) lattice of $L$ sites with open boundary conditions, reads 
\begin{equation}\label{eq:h}
\begin{aligned}
H =& \sum_{i=1}^{L-1} \Big[  -t\big (c^\dagger_i c_{i+1} + c^\dagger_{i+1} c_{i} \big )   + Vn_i n_{i+1}\Big] +\sum_{i=1}^{L}\epsilon_i  n_i\,.\\
\end{aligned}
\end{equation}
The hopping constant is set as the unit, $t=1$, throughout the paper. The onsite potentials are uniformly randomly distributed over the interval, $ \epsilon_i \in [-W/2, W/2]  $. The disorder strength is denoted by $W$ and the nearest-neighbor interaction strength by $V$.
The emergent picture, supported mostly by numerical studies \cite{Oganesyan2007,Pal2010,Luitz2015,BarLev2015,Bera2015,Mondragon-Shem2015}, is that a many-body localized
phase is stable at sufficiently strong disorder even in the middle of the many-body energy spectrum. As disorder strength is lowered, an ergodic phase 
is entered. The ergodic and MBL phase are distinct from each other in many ways: volume versus area-law scaling of the entanglement entropy \cite{Bauer2013,Kjaell2014},
thermalization versus failure of eigenstate thermalization hypothesis and Wigner-Dyson versus Poisson statistics of the level spacing distribution (see \cite{Nandkishore2015,Altman2015} for a review).
The reason for these behaviors is the existence of localized quasi-particles (often called l-bits) in the many-body localized phase \cite{Serbyn2013,Huse2014,Ros2015,Rademaker2016,Imbrie2016,Imbrie2016a} that form an extensive set of conserved quantities (see
\cite{Nandkishore2015,Rademaker2017,Imbrie2017} for a review).

Curiously, the conductivity does not behave according to the naive expectation (i.e., vanishing conductivities in the MBL phase and finite ones in the ergodic phase): in the MBL phase, numerical studies report vanishingly small dc conductivities,
consistent with a perfect insulator, while in the ergodic phase, some studies report finite dc-conductivities \cite{Barisic2016,Steinigeweg2016}, while others provide
evidence for subdiffusive dynamics \cite{BarLev2015,Agarwal2015,Gopalakrishnan2015,Luitz2016a,BarLev2017} (see also \cite{Bera2017} for a critical discussion and \cite{Luitz2017} for a review).

Most of the literature has, with few exceptions (see, e.g., \cite{Mondragon-Shem2015}), formally focussed  on repulsive interactions (see the discussion at the end of the introduction), while here we  explore the attractive regime.
In our work, we emphasize an  asymmetry in the phase diagram at weak disorder under sign changes of the interaction.
This asymmetry is traced back to the zero-temperature properties. While  local repulsive interactions cannot overcome disorder in the ground state which is thus fully localized, the situation is markedly different for attractive interactions. Bosonization studies \cite{giamarchi} and a subsequent
density matrix renormalization group analysis \cite{Schmitteckert1998} demonstrated the survival of a Luttinger liquid in the attractive case at half filling (see also \cite{Apel1982,Giamarchi1987,Giamarchi1988,Doty1992}).
This is not surprising since in general, disorder can even enlarge the region in parameter space for superfluid states at commensurate filling factors, as is particularly well known
for the case of bosons with repulsive interactions in the Bose-Hubbard model (see, e.g., \cite{Prokoviev1998,Rapsch1999,Gerster2016}).  
The reason are the disorder-induced density deviations (or, in other words, fluctuations in the local chemical potential) that drive density 
locally away from commensurability. 

\begin{figure}[t]
\includegraphics[width=\columnwidth]{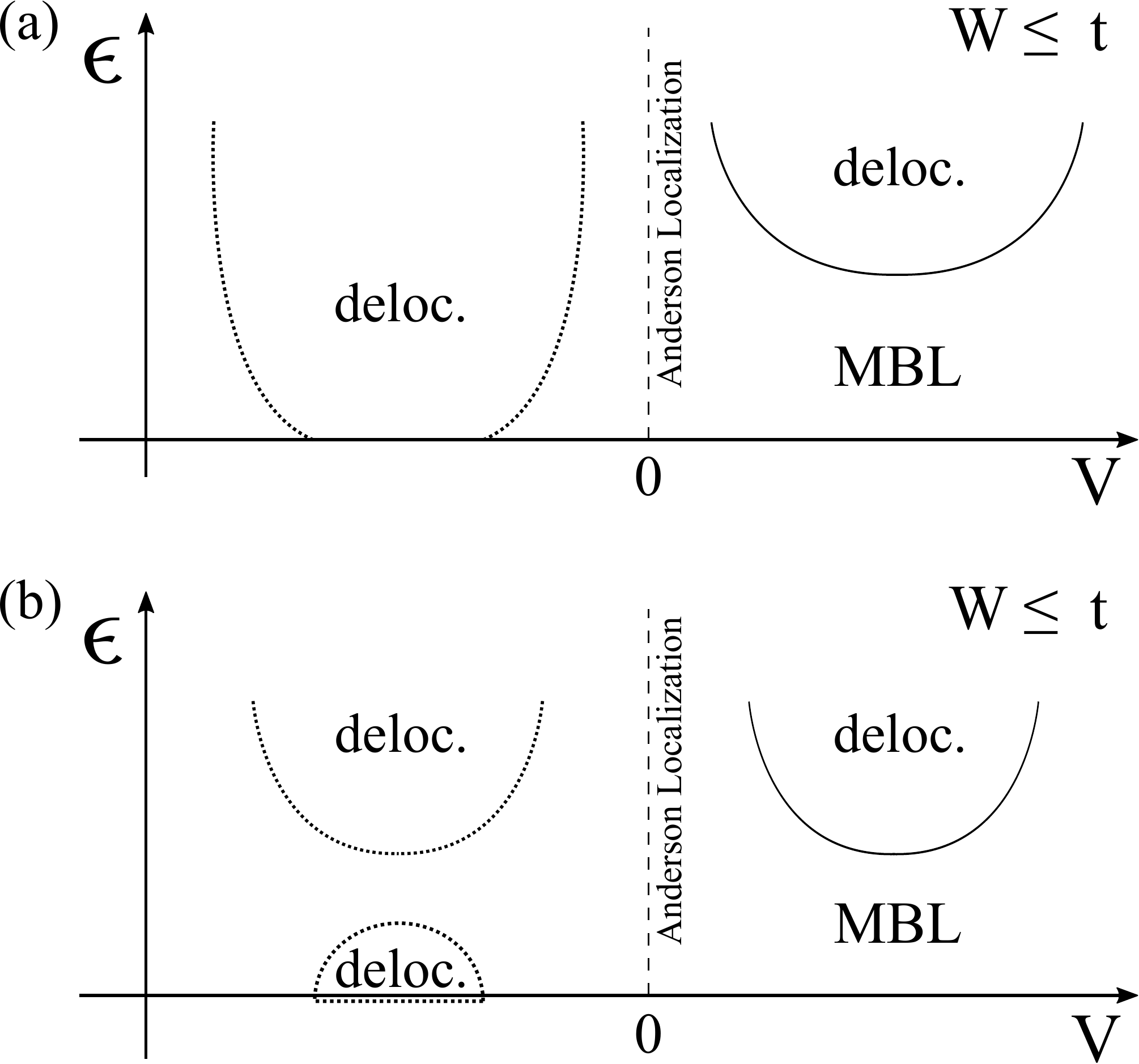}
\caption{Sketch of the energy density $\epsilon$ versus interaction strength $V$ phase diagram at weak disorder, with two  different scenarios.
(a) The Luttinger liquid is directly connected to an ergodic and delocalized phase at finite energy densities.
(b) The Luttinger liquid first transitions into the localized phase (either at an arbitrarily small energy density or at a finite energy density) 
 via an {\it inverted} mobility edge, while ultimately entering into the high-energy density delocalized  phase.
}
\label{fig:sketch}
\end{figure}

It is therefore an interesting question to ask how that zero-temperature Luttinger liquid connects to the finite-energy density phases \cite{Gornyi2005,Gornyi2007}.
Primarily, two scenarios are conceivable (see the sketch shown in Fig.~\ref{fig:sketch}). At large energy densities, similar to the case of
repulsive interactions, 
we expect an ergodic phase at weak disorder.
This high energy-density delocalized  phase can either be directly connected to the zero-temperature Luttinger liquid (the case shown in Fig.~\ref{fig:sketch}(a)) or it
could be intervened by a localized region (the case shown in Fig.~\ref{fig:sketch}(b)). The latter scenario
would imply an {\it inverted} mobility edge.

Such inverted mobility edges were suggested to exist in bosonic systems without disorder \cite{Pino2016}, and moreover,  even a recent experiment
with bosons in a two-dimensional lattice \cite{Choi2016} could indicate such a behavior, i.e., a delocalization-localization transition as energy density increases. 
In that experiment,
evidence for a many-body localized regime at high energy densities was found, while (minding details of the specific disorder distribution realized in that experiment),
theoretical studies predict a superfluid ground state \cite{Soyler2011}, and thus an extended state, for the values of interaction strength and disorder for which the experiment
reports localization.
Theoretically, there is evidence for the existence of an inverted mobility edge in the one-dimensional Bose-Hubbard model \cite{Singh2017,Sierant2017}. Ref.~\cite{Sierant2017} considered a two site
model with uncorrelated diagonal disorder while Ref.~\cite{Singh2017} arrives at the same conclusion yet there, the disorder is in the interactions. 
Clearly, additional studies are necessary to complete this question for the one-dimensional Bose-Hubbard model with disorder as well.

For our model of spinless fermions, we report evidence that the first scenario is realized, i.e., increasing energy density seems to immediately enlarge the 
delocalized region and there is no phase transition as energy density increases above the Luttinger-liquid phase. 
Our results are based on an analysis of the entanglement entropy \cite{Bauer2013,Kjaell2014}, the one-particle density matrix occupation spectrum \cite{Bera2015,Bera2017}, and the 
level-spacing distribution \cite{Oganesyan2007}.

An asymmetry between repulsive and attractive interactions has also been found in \cite{Mondragon-Shem2015},
where the limit of strong interactions in a system of spinless fermions with disorder was considered. In the equivalent language of spin-1/2 degrees
of freedom, this corresponds to analyzing the effect of disorder on  states with ferromagnetic versus antiferromagnetic order at low energies. In
the ground state, both phases become localized upon adding disorder. The strongly attractive
side was found to lead to a more stable many-body localized phase at finite energy densities compared to the antiferromagnetic case, which 
shifts the mobility-edge to higher temperatures.

We should stress that while formally, most studies focussed on repulsive interactions, one can use particle-hole symmetry 
(which is violated in individual disorder realizations but restored after disorder averaging)
to relate the negative $V$ side to the positive $V$ side (see the discussion in \cite{Naldesi2016}).
As a consequence, the low-energy density region at $V<0$ maps to the high-energy density at the $V>0$ side.
Using this argument and by inspection of existing studies of the  energy-density versus disorder 
phase diagram (see, e.g., \cite{Mondragon-Shem2015,Bera2017,Nag2017,Luitz2015,Naldesi2016}),
one can already draw some conclusions on the structure of the phase diagram on the attractive side.
The special point of $V=2t$ is the most studied one, for which there are also full  energy-density versus disorder strength
phase diagrams available (see, e.g., \cite{Bera2017}).
For the purpose of clarifying the question of an inverted mobility edge on the attractive side,
no conclusive picture arises from the existing results for $V=2t$: according to Refs.~\cite{Schmitteckert1998,Doggen2017}, $V=-2t$ sits right
at the edge of the Luttinger-liquid phase and any finite disorder drives the system into the localized regime
in the ground state (see, e.g., the ground-state phase diagram presented in Figs.~\ref{fig:contour} and \ref{fig:FRG_data}(a)).
Therefore, in order to answer the question of an inverted mobility edge in this model and to complete the analysis
of its full interaction strength - disorder phase diagram, a study of the behavior at weak interaction $0<V < -2t$ and disorder strengths $0<W < 2t$ is necessary.

This paper is organized as follows. In Sec.~\ref{sec:model}, we define the model and the quantities computed in this work. In Sec.~\ref{sec:gs}, we revisit
the ground-state phase diagram and show that the occupation-spectrum discontinuity is smaller in the Luttinger-liquid phase, while finite-size dependencies
prohibit an extraction of the phase boundaries from that quantity. We further demonstrate that 
 the zero-temperature disorder-interaction phase diagram can also be obtained from a calculation of the conductance using the functional renormalization group method, with qualitative agreement with other methods.  Section~\ref{sec:finite-e} contains our main results for the phase diagrams 
at finite energy density and attractive interactions, obtained from exact diagonalization. We conclude with Sec.~\ref{sec:sum}, where we also discuss open questions.


\section{Model and observables,  and numerical methods}
\label{sec:model}

We consider the spinless-fermion model defined in Eq.~\eqref{eq:h}.
Our  choice of units is the one employed in \cite{Schmitteckert1998}, while  in most of the current MBL literature, the choice of units derives from the 
equivalent formulation of Eq.~\eqref{eq:h} in the spin language \cite{Pal2010,Luitz2015}, where then $\epsilon \in [-W', W']$ and $t\to t'/2$. Hence, in order to compare with 
the units used in that literature, our numbers have to be divided by a factor of 4. For instance, the critical value for the delocalization-localization transition
for a Heisenberg chain ($V=2t$ in our units) is, at $T=\infty$, given by $W'\approx 3.5 t'$ \cite{Luitz2015}, hence in our units, at $W\approx 14t$.

We further consider the half-filling subspace $N=L/2$ and obtain results for both attractive  and repulsive interactions. The target energy density is defined as $\epsilon = 2(E_{\rm max}-E)/(E_{\rm max}-E_{\rm min})$. Note that a different convention for $\epsilon$ is used in \cite{Luitz2015,Mondragon-Shem2015,Naldesi2016}.  The arithmetic average over disorder realizations is denoted as $[\ \cdot\ ]$.

The observables calculated are the following ones. The one-particle density matrix (OPDM) is defined as 
\begin{equation}
\rho_{ij} = \braket{ \psi | c^\dagger_i c_j | \psi}  
\end{equation}
where $\ket{\psi}$ is the given many-body wavefunction. We here compute the OPDM in many-body eigenstates $|\psi\rangle = |n\rangle$ with $H|n\rangle=E_n|n\rangle$. By diagonalizing the OPDM,
\begin{equation}
\rho \ket{ \phi_\alpha } = n_\alpha \ket{\phi_\alpha}
\end{equation}
one obtains the occupation spectrum $n_\alpha$ and the natural orbitals $\ket{\phi_\alpha}$, which form a complete basis set of single-particle states for each $|n\rangle$. After ordering the occupation spectrum, i.e.,  $n_1\geq n_2 \geq \dots \geq n_L$, the discontinuity in the occupation spectrum is defined by $\Delta n = n_{L/2-1} - n_{L/2}$
since we have $N=L/2$ particles in the system with $L$ even.

In a noninteracting fermionic system, any many-body wave function can be written as a Slater determinant of single-particle states. In this case, there will be a set of the natural orbitals with $n_\alpha = 1$, spanning the same vector space as the single-particle eigenstates and therefore, 
the occupation spectrum is a step function (after reordering the eigenvalues \cite{Bera2017}).
 As a result, a discontinuity $\Delta n=1$ in the occupation spectrum  corresponds to product states in Fock space. One refers to this also as localization in Fock space and a large discontinuity 
with many eigenvalues being close to one or zero implies only a very weak Fock-space delocalization, reminiscent of a zero-temperature Fermi-liquid \cite{Bera2017}.
The MBL phase has a nonzero occupation-spectrum discontinuity in our model with repulsive interactions, as was shown in \cite{Bera2015,Bera2017} and also
in a system of hardcore bosons in two dimensions \cite{Inglis2016}. The properties of OPDMs in systems of one-dimensional hardcore bosons on a lattice and in the presence of disorder
was studied in \cite{Nessi2011,Gramsch2012}. 
Note that the occupation-spectrum discontinuity is smeared out in extensive superpositions of many-body eigenstates even in the MBL phase, while the  distribution as such remains
highly nonthermal (see the example of a quantum quench discussed in \cite{Lezama2017}). 

  To determine whether particles in the MBL phase are  truly localized in real space, we also calculate the inverse participation ratio (IPR) of the natural orbitals, which we define as
\begin{equation}
\text{IPR} = \frac{1}{L}\sum_{\alpha=1}^L \sum_{i=1}^L | \phi_\alpha(i) |^4\,. 
\label{eq:ipr}
\end{equation}
The inverse participation ratio unveils the localization in real space by taking into account the spatial structure of the natural orbitals. We see that in the limiting cases, $\text{IPR} = 1/L$ in the ergodic phase and $\text{IPR} = 1$ in the localized phase.
Note that unlike in \cite{Bera2015}, we do not weigh the natural orbitals by their occupations. The analysis of the statistical properties of the natural
orbitals across the entire spectrum for a single disorder realization presented in \cite{Bera2017} shows that the 
natural orbitals are, in first approximation, very similar from many-body to many-body eigenstate, implying that 
mostly just the occupations change. 

The von-Neumann entropy is the bipartite entanglement entropy defined as $S_{vN} = -\mbox{Tr} \lbrack \rho_A \ln \rho_A \rbrack$, 
where $\rho_A$ is the reduced density matrix obtained by tracing out half of the system. In the localized regime, the entanglement entropy obeys an area-law scaling, while in the ergodic regime, a volume-law scaling is found \cite{Bauer2013}. The variance of the entanglement entropy is expected to diverge at the transition \cite{Kjaell2014}.

The many-body eigenenergy spectra in the localized and ergodic regimes have different distributions \cite{Oganesyan2007}. Therefore, a standard way to study many-body localization is to analyze the different statistics of the adjacent level spacing \cite{Oganesyan2007,Serbyn2016}. The adjacent gap ratio is defined as 
\begin{equation} r_{\rm gap} = {\rm min}(\delta^{(n)},\delta^{(n+1)})/{\rm max}(\delta^{(n)},\delta^{(n+1)})\,,
\end{equation} 
where $\delta^{(n)}=E_n - E_{n-1}$ is the difference of adjacent many-body energy levels. 
Random matrix theory predicts that the distribution of level spacings themselves 
follows  a Poisson distribution in the localized regime and a Gaussian orthogonal ensemble  in the ergodic phase.
This implies certain values for the gap ratio in these two cases, namely $r_{\rm gap}=0.3863$ and $r_{\rm gap}= 0.5307$, respectively (see the discussion in \cite{Oganesyan2007}).

 \subsection{Density matrix renormalization group}
In order to compute the occupation-spectrum discontinuity in the ground state, we use density  matrix renormalization group simulations \cite{White1992,Schollwoeck2005} in a 
matrix-product state representation using a single-site algorithm with subspace expansion \cite{Hubig2015}. 
We use up to 400 states leading to discarded weights of $10^{-14}$ and an average over 5000 disorder realizations.

\subsection{Exact diagonalization}
\label{sec:methods}
In order to map out the phase diagram at finite energy densities, 
we study the model by exact diagonalization for  system sizes $L=12,14,16,18$ with at least $3000$ realizations for each parameter chosen. The exact diagonalization study of many-body localization is limited by both the large dimension of the subspace, e.g., $dim(H_{18})=48620$ and the requirement for a sufficient number of realizations. For all quantities calculated, except for the adjacent gap ratio, we take only the closest eigenvector for the target energy density in each realization.  However, we find it necessary to take the 50 closest eigenvalues in one realization to have acceptable statistics for the  adjacent gap ratio. We study the model with up to $L=18$ sites at various energy densities $\epsilon = \{ 0.025, 0.05, \dots, 1\}$ with the shift-invert spectral transformation $H\rightarrow (H - \lambda I )^{-1}$, where $\lambda$ is the targeted eigenvalue. 
The libraries PETSc and SLEPc \cite{Hernandez:2005:SSF} are used for the problem setup and building the solution. We use an exact shift-and-invert solver. The inverse problem involved is solved by direct linear solver with external package MUMPS \cite{amestoy2001fully,amestoy2006hybrid} using parallel Cholesky factorization. 

\begin{figure}[t]
\includegraphics[height=1.0cm,clip]{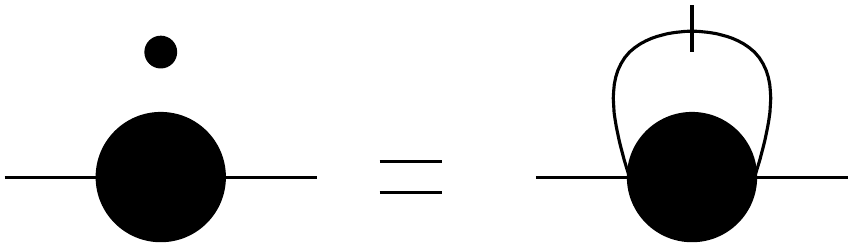}\\
\includegraphics[height=1.0cm,clip]{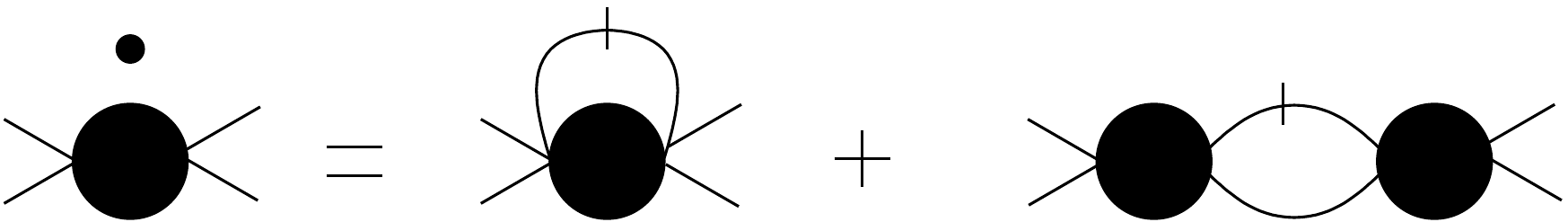}
\caption{(Color online) Schematic representation of the flow equations for the self-energy and the effective two-particle scattering (an $n$-particle vertex has $2n$ external legs).  }
\label{fig:floweq}
\end{figure}

\subsection{Functional Renormalization Group}

The functional renormalization group (FRG) is one implementation of Wilson's general RG idea for interacting many-particle systems \cite{frgrev2}. It can be set up either on the Keldysh contour or the Matsubara axis; since we are interested in ground-state properties, the latter choice is the more convenient one. The starting point of an FRG calculation is to take the noninteracting Green's function $G^0$ of the system under consideration and to cut it off below an infrared energy scale $\la$. In particular, we introduce a multiplicative cutoff in Matsubara frequency space,
\begin{equation}\label{eq:cutoff}
 G^{0,\la}(i\omega) = \Theta(|i\omega|-\la) G^0(i\omega)\,,
\end{equation}
and consider the flow of many-particle vertex functions (such as the self-energy or the effective interaction) as a function of $\la$. The resulting (infinite) set of coupled flow equations can be represented elegantly using Feynman-like diagrams (see Fig.~\ref{fig:floweq}). Subsequent re-integration from $\la=\infty$ down to the cutoff-free system $\la=0$ amounts to an exact solution of the many-particle problem. In practice, the infinite hierarchy of flow equations needs to be truncated, rendering the FRG an approximate method which treats interactions perturbatively.

The simplest truncation scheme is to only consider the flow of the single-particle vertex (the self-energy) and to neglect the flow of all higher-order vertex functions. If the two-particle vertex is set to the bare interaction $V$, the flow of the self-energy associated with Eq.~(\ref{eq:h}) can be expressed simply in terms of effective hopping-matrix elements $t_l^\la$ and on-site energies $\epsilon_l^\la$:
\begin{eqnarray}
\partial_\la \epsilon_l^\la &=& -\frac{1}{\pi}\tn{Re}\,\big[V_{l-1} \tilde G_{l-1,l-1}^\la(i\la)+V_{l} \tilde G_{l+1,l+1}^\la(i\la)\big]\,, \nonumber \\
\partial_\la t_l^\la & =& -\frac{1}{\pi}\tn{Re}\,\big[V_l \tilde G_{l,l+1}^\la(i\la)\big]\,,
\label{eq:flow}
\end{eqnarray}
where $\tilde G^\la(i\omega)$ denotes the flowing single-particle Matsubara Green's function (an explicit expression can be found in Eq.~(\ref{eq:gf})). At half filling, the initial conditions are given by $\epsilon_l^{\la\to\infty}=\epsilon_l$ as well as $t_l^{\la\to\infty}=t$. We explicitly allow for a spatial dependence of the two-particle interaction $V_l$ in order to model a smooth coupling to free fermion source and drain leads (see below). Boundary conditions are formally imposed by setting $V_{-1}=V_{L}=0$.

The approximation introduced above is strictly correct only to leading order in the interaction $V$ but contains an infinite resummation of Feynman diagrams (since the self-energy feeds back into its own flow). Similarly, one can obtain a second-order truncation scheme by accounting for the flow of both the two-particle vertex and the self-energy while setting all higher-order vertices to their initial value (zero). In this paper, we partially incorporate second-order contributions by parameterizing the two-particle vertex in terms of flowing effective onsite interactions $V_l^\la$. This is a purely pragmatic approach -- the resulting approximation is still strictly controlled only to first order -- which improves our results quantitatively but does not change them qualitatively. The corresponding set of flow equations is given by Eq.~(\ref{eq:flow}) with $V_l\to V_l^\la$ and complemented by a flow equation for the effective interaction $V_l^\la$, which we do not write down explicitly; it can be found in Ref.~\onlinecite{LLdisorder}.

In order to compute the conductance, we couple the system to left and right (source and drain) leads, which, for reasons of simplicity, we model as structureless Fermi liquids (e.g., we take the wide-band limit). Such a free-fermion system can be  `projected out' analytically via equation-of-motion techniques and the calculation of $\tilde G^\la(i\omega)$ reduces to the inversion of a $L\times L$ matrix defined by 
\begin{equation}\label{eq:gf}\begin{split}
& \big[\tilde G^\la(i\omega)^{-1}\big]_{l,l} = i\omega - \epsilon_l^\la +i\Gamma\sgn(\omega)(\delta_{l,1}+\delta_{l,L})\,, \\
& \big[\tilde G^\la(i\omega)^{-1}\big]_{l,l+1} = \big[\tilde G^\la(i\omega)^{-1}\big]_{l+1,l} = t_l^\la\,.
\end{split}\end{equation}
Due to the tridiagonal structure, this inversion can be carried out with a computational effort scaling linearly with $L$. The flow equations (\ref{eq:flow}) can be integrated using standard Runge-Kutta routines. Finally, one obtains the conductance $g$ (in units of $e^2/h=1$) from
\begin{equation}\label{eq:cond}\begin{split}
g(L) = 4\Gamma^2 \big|\tilde G^{\la=0}_{1,L}(i\omega\to \omega + i0)\big|^2\,.
\end{split}\end{equation}
More details on the FRG can be found, e.g., in Refs.~\onlinecite{frgrev2,LLdisorder}.


\section{Ground-state properties}
\label{sec:gs}

The existence of a Luttinger-liquid phase in the presence of disorder in the spin-1/2 XXZ chain is a much studied problem,
which is equivalent to spinless fermions or hardcore bosons. While early work established the existence of such a phase \cite{Apel1982,Giamarchi1987,Giamarchi1988,Doty1992,Schmitteckert1998,Schuster2002,Urba2003,Carter2005},
there is an ongoing discussion on the nature of the transition between the delocalized superfluid and the localized phase (which, in the language of bosons, is 
a Bose-glass phase \cite{Fisher1989}). This question is not at the focus of our work and we refer the reader to the pertinent literature for details \cite{Altman2004,Altman2008,Altman2010,Ristivojevic2012,Ristivojevic2014,Pielawa2013,Pollet2013,Pollet2014,Yao2016,Doggen2017}.

\begin{figure}[!t]
\includegraphics[width=\columnwidth]{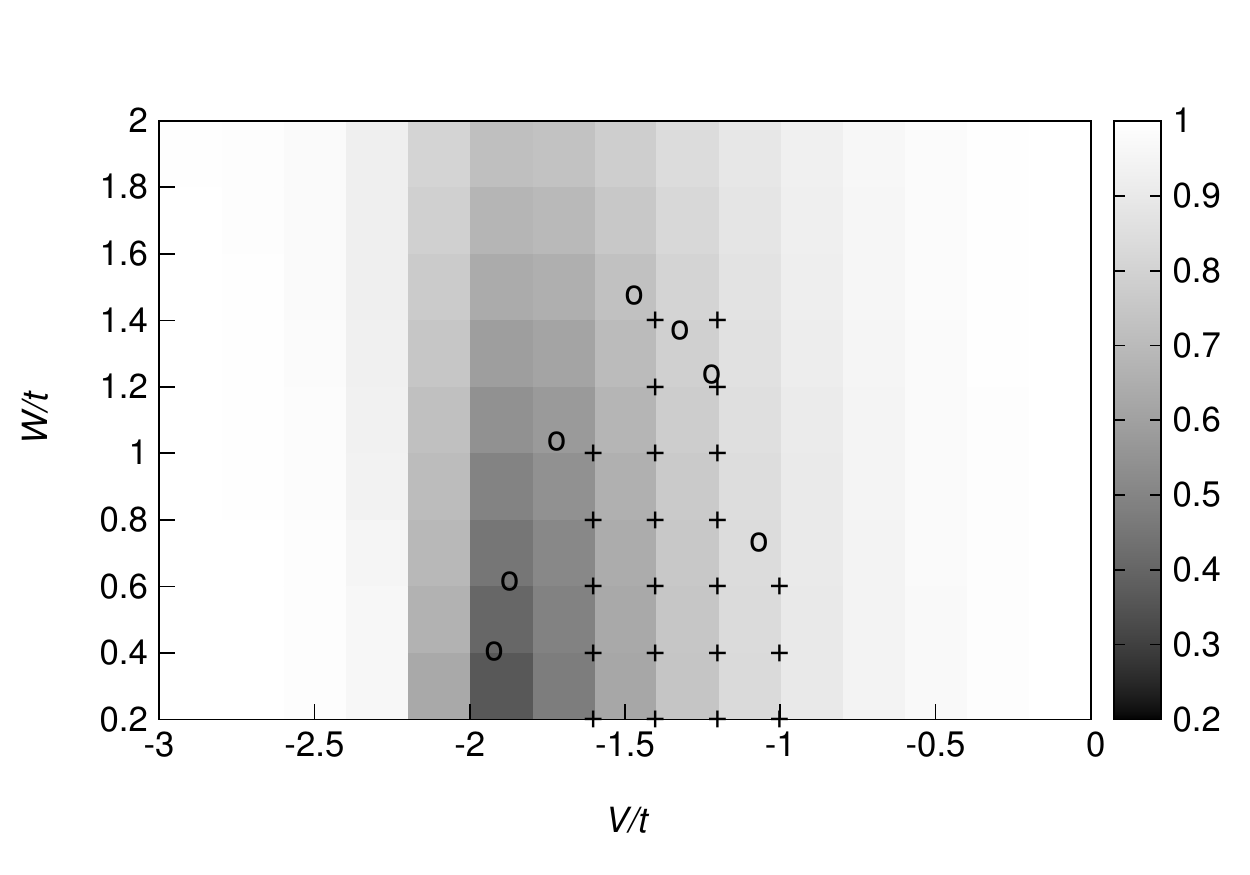}
\caption{(Color online)
Occupation-spectrum discontinuity $\lbrack \Delta n \rbrack$ (grey-shaded) in the ground state compared to results from \cite{Schmitteckert1998} ($+$ signs)
as a function of the interaction strength $V$ and disorder $W$ for  $L=32$ (DMRG data).
A clear feature emerges in this quantity in the vicinity of the region where the system
is in a delocalized phase. The crosses are the results from \cite{Schmitteckert1998}; circles are the more recent phase boundaries
from \cite{Doggen2017}.
}
\label{fig:contour}
\end{figure}

\subsection{Occupation-spectrum discontinuity}

An early DMRG study used the sensitivity to twisted boundary conditions to locate the transition between localized and extended phase. \cite{Schmitteckert1998}.
Entanglement measures were used  to locate this transition in \cite{Zhao2013,Berkovits2015} and it was pointed out \cite{Berkovits2015} that large system sizes
are needed to be in the correct scaling regime, due to the large localization length.

Here, we show the phase diagram from Ref.~\cite{Schmitteckert1998} in the disorder-interaction strength plane in Fig.~\ref{fig:contour}. 
The figure shows the value of the occupation-spectrum discontinuity $\lbrack \Delta n\rbrack $ computed for $L=32$ sites using DMRG simulations. 
Remarkably, the region
where $\lbrack \Delta n\rbrack $ is small (compared to large $W$ or small $|V|/J$) coincides with the Luttinger liquid (indicated by the symbols).
We attempted an extrapolation of $\lbrack \Delta n\rbrack $ in the  system size to locate the transition yet obtained no conclusive results, which we attribute
to the small system sizes $L\leq 128$ considered.

\begin{figure}[!t]
\includegraphics[width=\columnwidth]{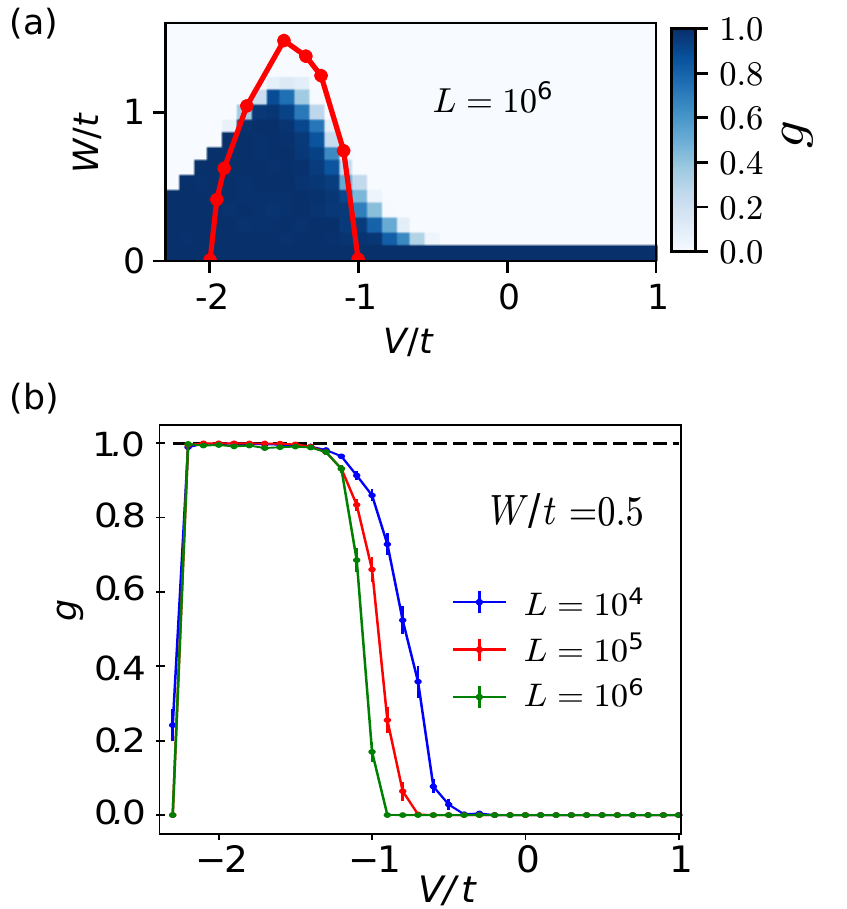}
\caption{(Color online)
(a) Conductance $g(L=10^6)$ as a function of the interaction strength $V$ and disorder $W$ at zero temperature as computed from the functional renormalization group. For comparison, the phase boundary from Ref. \cite{Doggen2017} is shown via the filled circles.  The FRG data represents a disorder average over $30$ samples per data point. (b) Cut through the phase diagram at $W/t=0.5$ for various system sizes $L$.}
\label{fig:FRG_data}
\end{figure}

\subsection{Conductance from FRG}

It is known \cite{giamarchi}, that for weak disorder the localization length diverges as $\xi\sim W^{-2/(3-2K)}$ as one approaches $V=-t$ from above (where the Luttinger-liquid parameters takes the value $K=3/2$). This motivates us to study the problem using a different method -- the FRG -- whose strengths and shortcomings are orthogonal to those of an exact-diagonalization approach.

The key disadvantage of the FRG is that it is approximate w.r.t.~the two-particle interaction $V$. Within our truncation scheme, all results are strictly correct only to leading order in $V$; higher-order contributions are uncontrolled. Thus, the FRG is not a suitable tool to determine the precise position of a phase boundary which is situated away from $V/t\ll 1$. The advantage of the FRG is that one can easily treat large systems of $L=10^6$ sites and that leads as well as single-particle disorder can be incorporated exactly. Hence, the FRG can overcome the obstacle of a potentially large localization length \cite{Berkovits2015} and provide further evidence for the existence of the delocalized phase in certain parameter regimes.

We use the FRG to compute the disorder-averaged conductance $g$ in the ground state. To this end, one couples the system to left and right Fermi-liquid leads; in order to avoid backscattering at the interface between the noninteracting leads and the interacting region, we employ a spatially smoothened transition region of about 20 sites. The metallic and localized phases are then characterized by $g\sim\textnormal{const.}$ and $g\sim e^{-L}$ at large $L$, respectively. In Fig.~\ref{fig:FRG_data}(a), we show $g$ in the $V-W$ parameter plane for a fixed large system size $L=10^6$, which qualitatively reproduces Fig.~\ref{fig:contour}. The interaction-dependence of the conductance for a fixed disorder strength of $W/t=0.5$ is displayed in Fig.~\ref{fig:FRG_data}(b) for three different system sizes ranging from $L=10^4$ to $L=10^6$. This illustrates that $g$ becomes independent of $L$ for $-2t\lesssim V\lesssim -t$ and thus provides further evidence for the existence of the metallic phase in this regime. Note that our FRG scheme apparently misses the localization at large attractive $V\sim -2t$ (see below).

It is surprising that our simple-minded `Hartree-Fock-like' FRG approach captures the transition between the delocalized and the metallic phase despite the fact that this transition does not occur for $|V|/t\ll1$ where the approximation is controlled. This observation motivates an extension of the FRG to a second-order scheme, which would allow one to access finite-energy properties for systems of up to $L=10^2$ sites. Such an investigation is currently under way; preliminary results show that one can capture the phase-separation transition for large negative $V$ that is missed by the leading-order scheme.

\section{Finite energy densities}
\label{sec:finite-e}

We now turn to our main case of interest, finite energy densities $\epsilon>0$.
Our analysis of the energy-density versus interaction-strength phase diagram is based on
calculating the impurity averaged one-particle density matrix occupation-spectrum discontinuity $\lbrack \Delta n \rbrack$ \cite{Bera2015}, the (half-cut) von-Neumann
entropy $\lbrack S_{\rm vN} \rbrack$ \cite{Bauer2013}, its variance $\mbox{var}\,S_{\rm vN}$ \cite{Kjaell2014} and the level-spacing distribution \cite{Oganesyan2007,Pal2010}.
Our main goal is to illustrate the quantitative and qualitative differences between attractive and repulsive interactions
at weak disorder and low energy densities.

\subsection{Occupation-spectrum discontinuity}

\begin{figure}[t]
\includegraphics[width=\columnwidth]{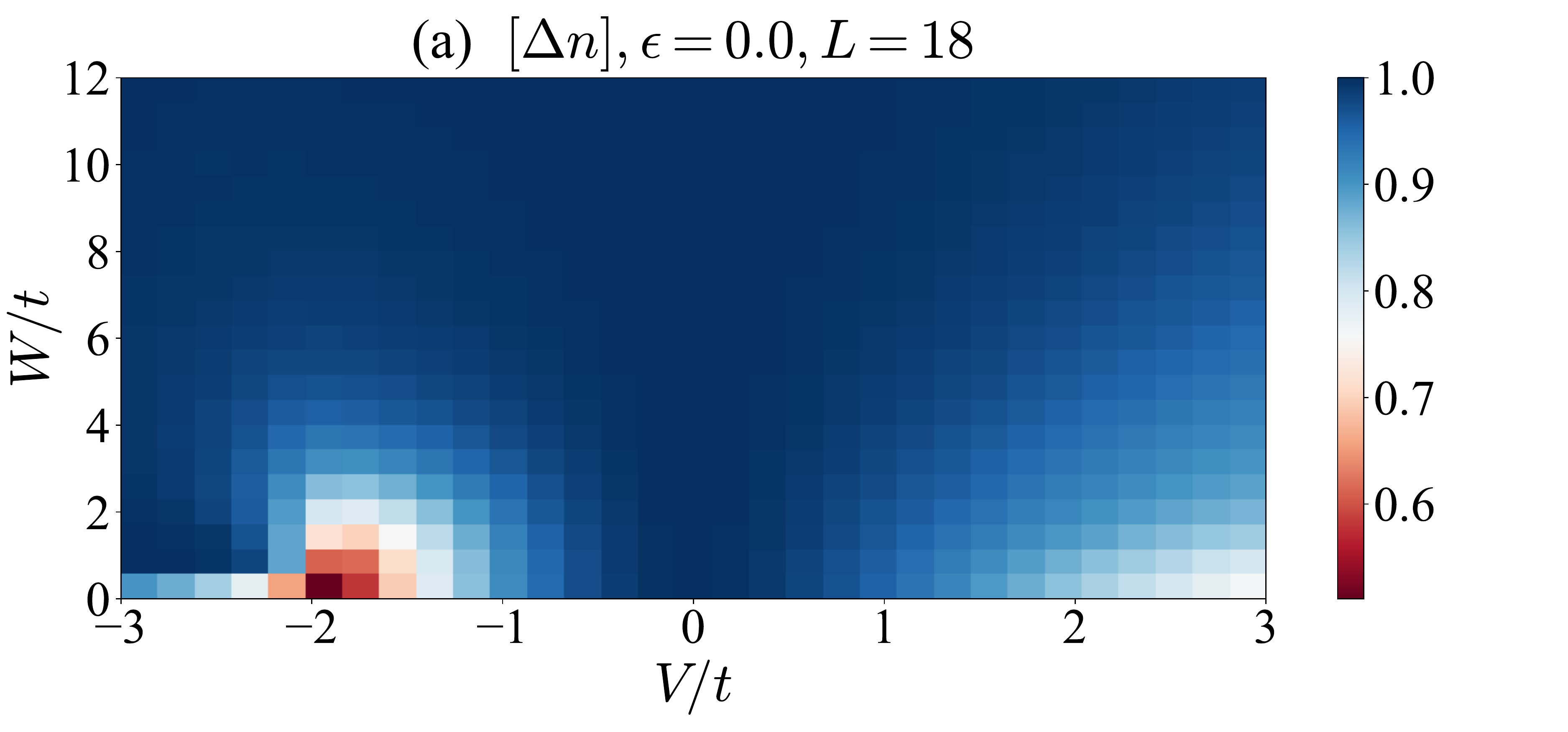}
\includegraphics[width=\columnwidth]{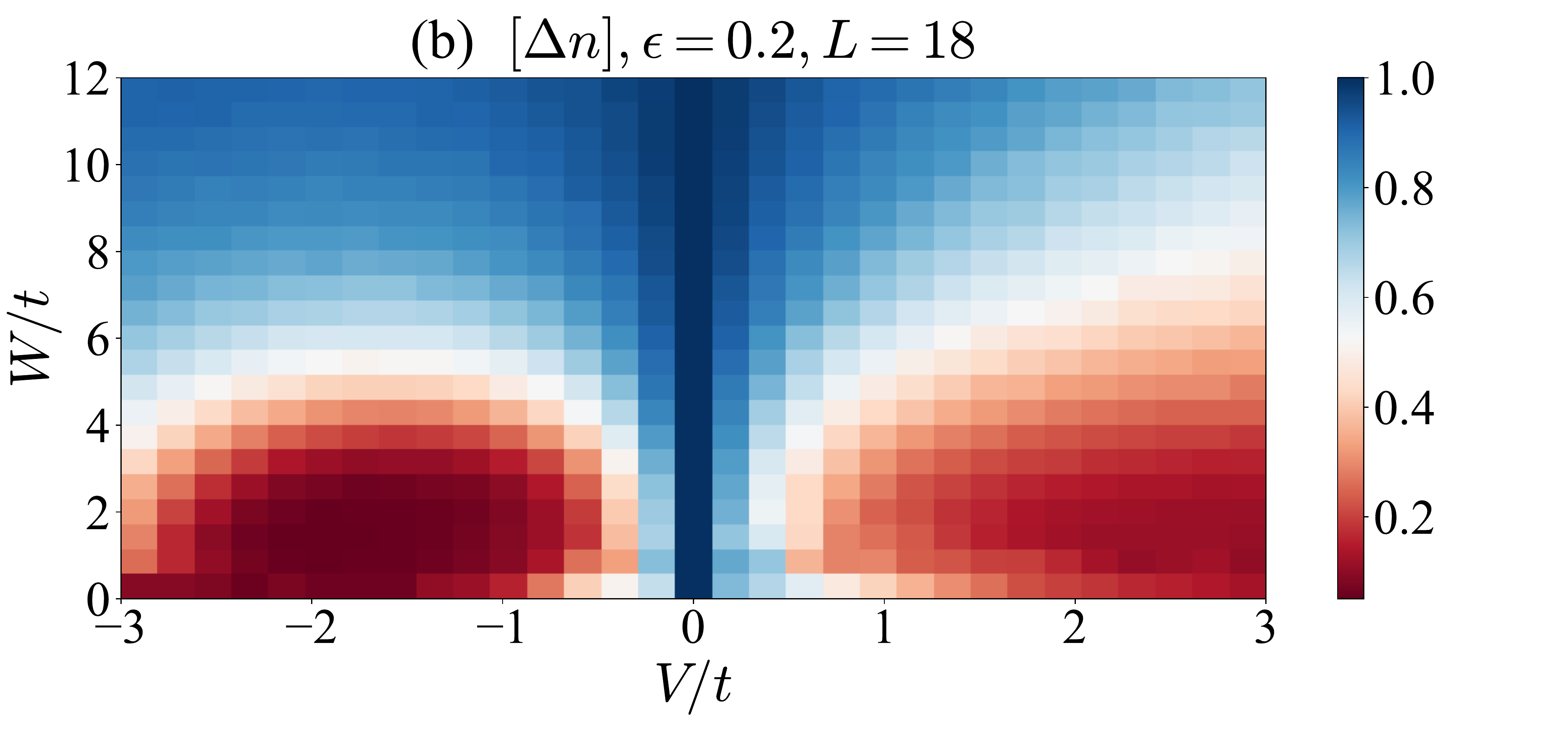}
\includegraphics[width=\columnwidth]{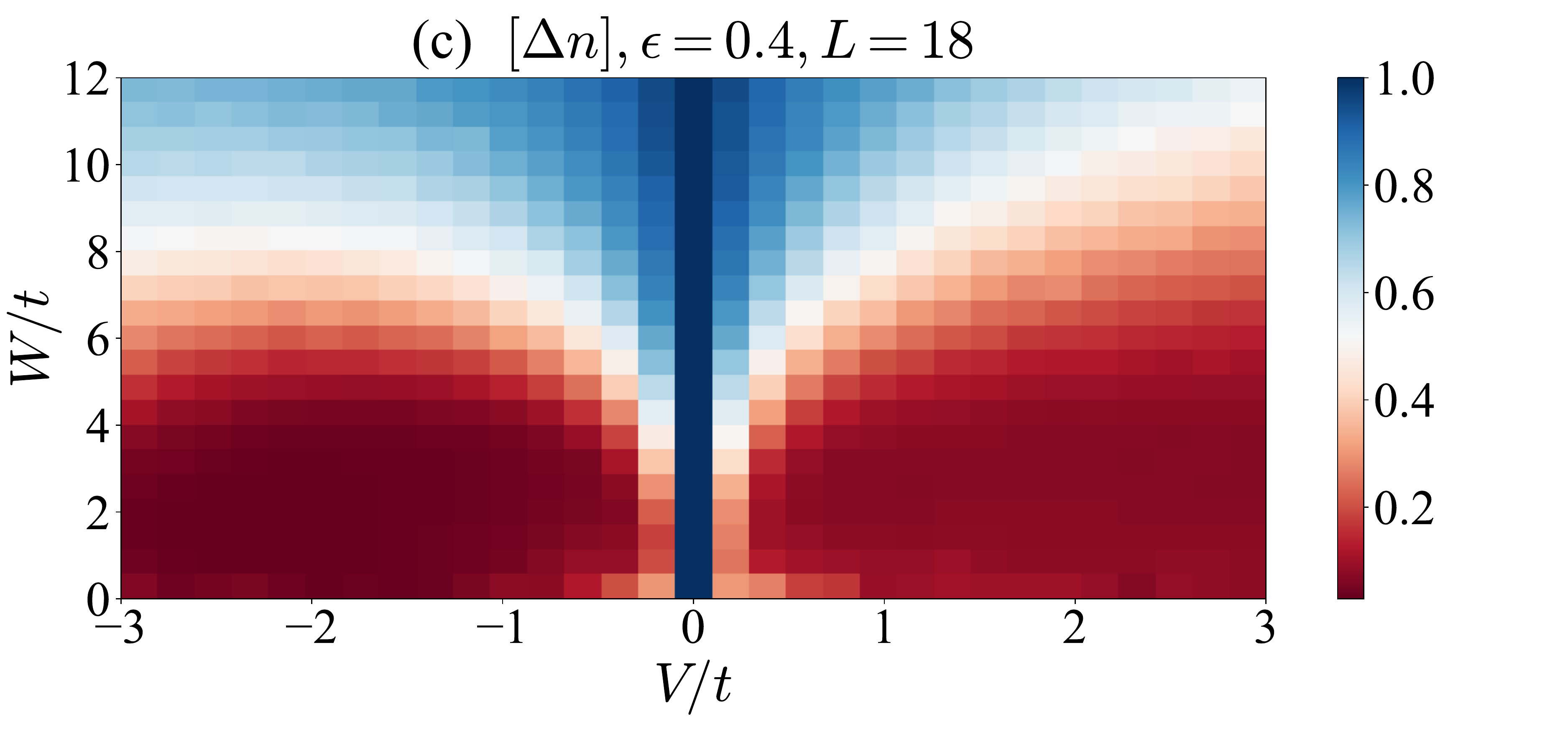}
\includegraphics[width=\columnwidth]{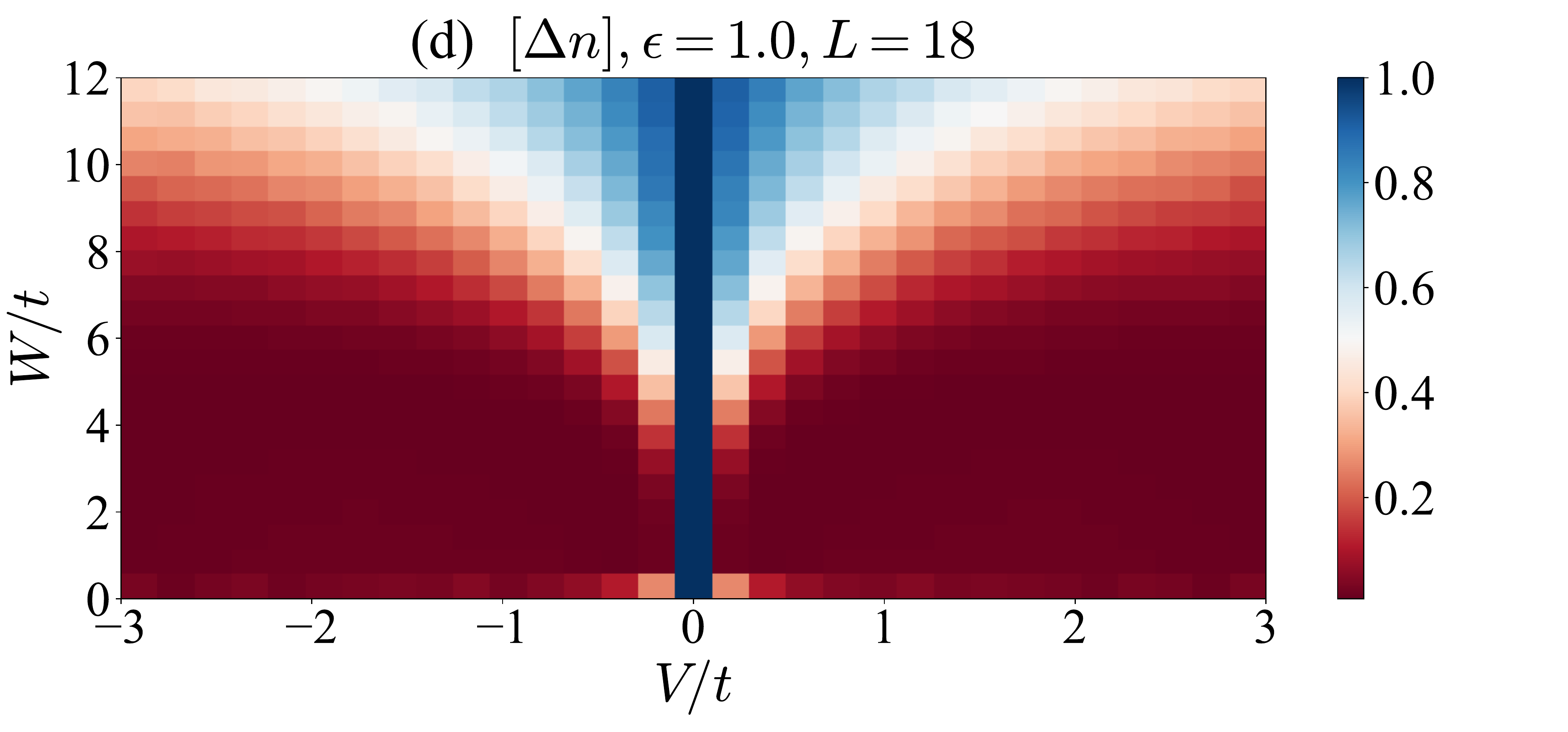}
\caption{OPDM occupation discontinuity in the $W$ versus $V$ plane for (a) $\epsilon=0 $, (b) $\epsilon=0.2 $, (c) $\epsilon=0.4 $, (d) $\epsilon=1 $ for $L=18$.
Note the different scale for the color coding used in (a).
}
\label{fig:deltan_W_V}
\end{figure}

The occupation-spectrum discontinuity $\lbrack \Delta n\rbrack$ is plotted in Figs.~\ref{fig:deltan_W_V}(a)-(d)
in the disorder-versus-interaction strength plane for different values of the energy density $\epsilon=0, 0.2, 0.4, 1$. 
Note that the occupation-spectrum discontinuity at $V=2t$ was already computed for all energy densities in \cite{Bera2017}.

Figure~\ref{fig:deltan_W_V}(a) qualitatively reproduces, albeit for much smaller systems, the behavior in the ground state discussed in Sec.~\ref{sec:gs}:
in the vicinity of the Luttinger-liquid phase at $-2 \lesssim V/t \lesssim -1.5$, the occupation-spectrum discontinuity 
is markedly smaller than anywhere else in the phase diagram and the $V\to -V $ asymmetry is apparent.

Upon increasing energy density, at both $V>0$ and $V<0$, a region with small $\lbrack \Delta n\rbrack$
emerges at small values of the disorder strength and quickly grows in size. For negative $V<0$, this presumably ergodic region 
appears to be adiabatically connected to the 
zero energy-density Luttinger liquid, as suggested by the sequence of results for increasing $\epsilon$ presented in Figs.~\ref{fig:deltan_W_V}(a)-(d).
There is also a regime with a reduced occupation-spectrum discontinuity at $V\sim 3t$ on the repulsive side and weak disorder. This one, 
however, does not become smaller as system size increases, contrary to the behavior in the Luttinger-liquid regime at negative values of $V$ (data not shown here).
We speculate that this reduction in $\Delta n$ for large $V$ and small disorder is inherited from the ground-state degeneracy at $W=0$ and $V\gg t$
in the density-wave phase.

\begin{figure}[t]
\includegraphics[width=\columnwidth]{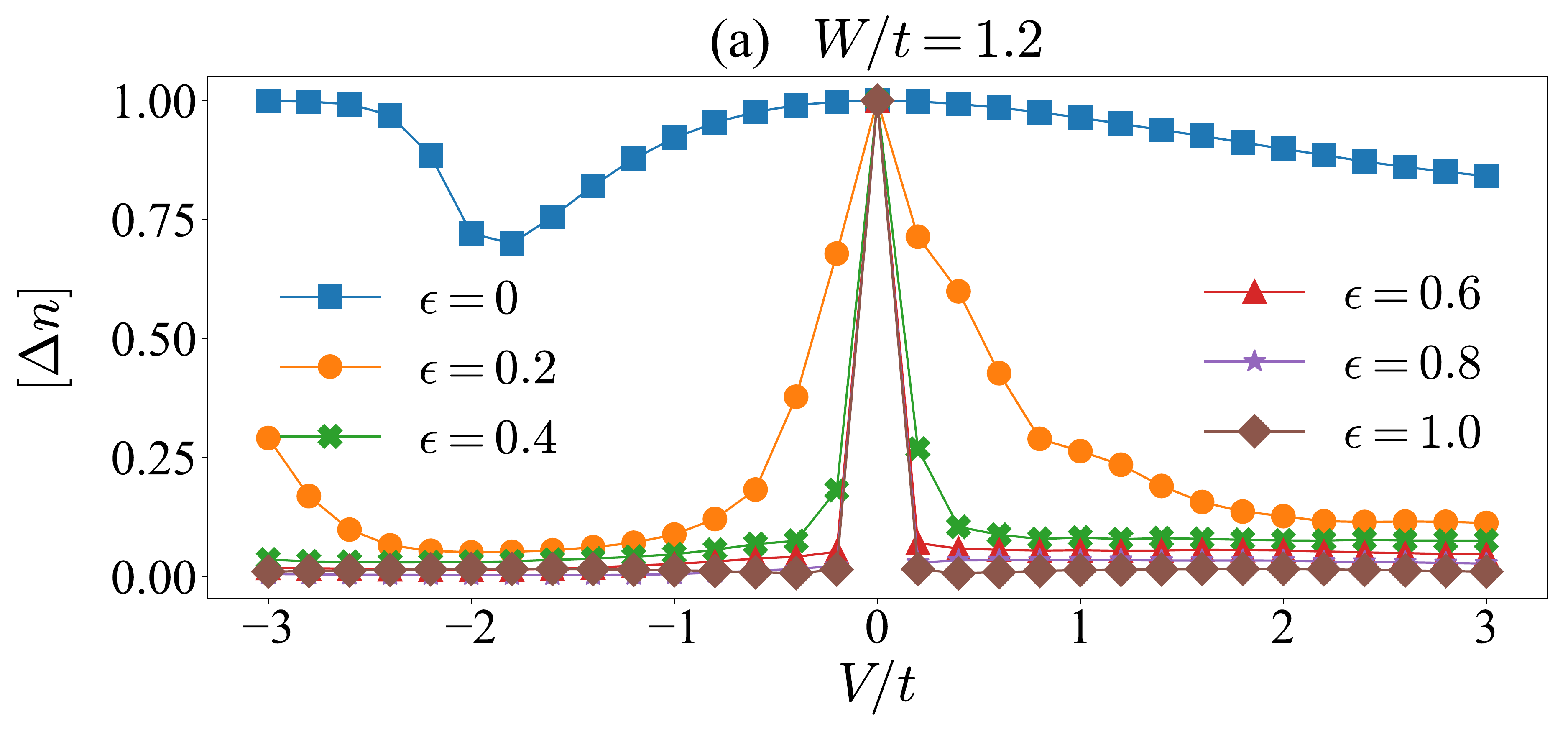}
\includegraphics[width=\columnwidth]{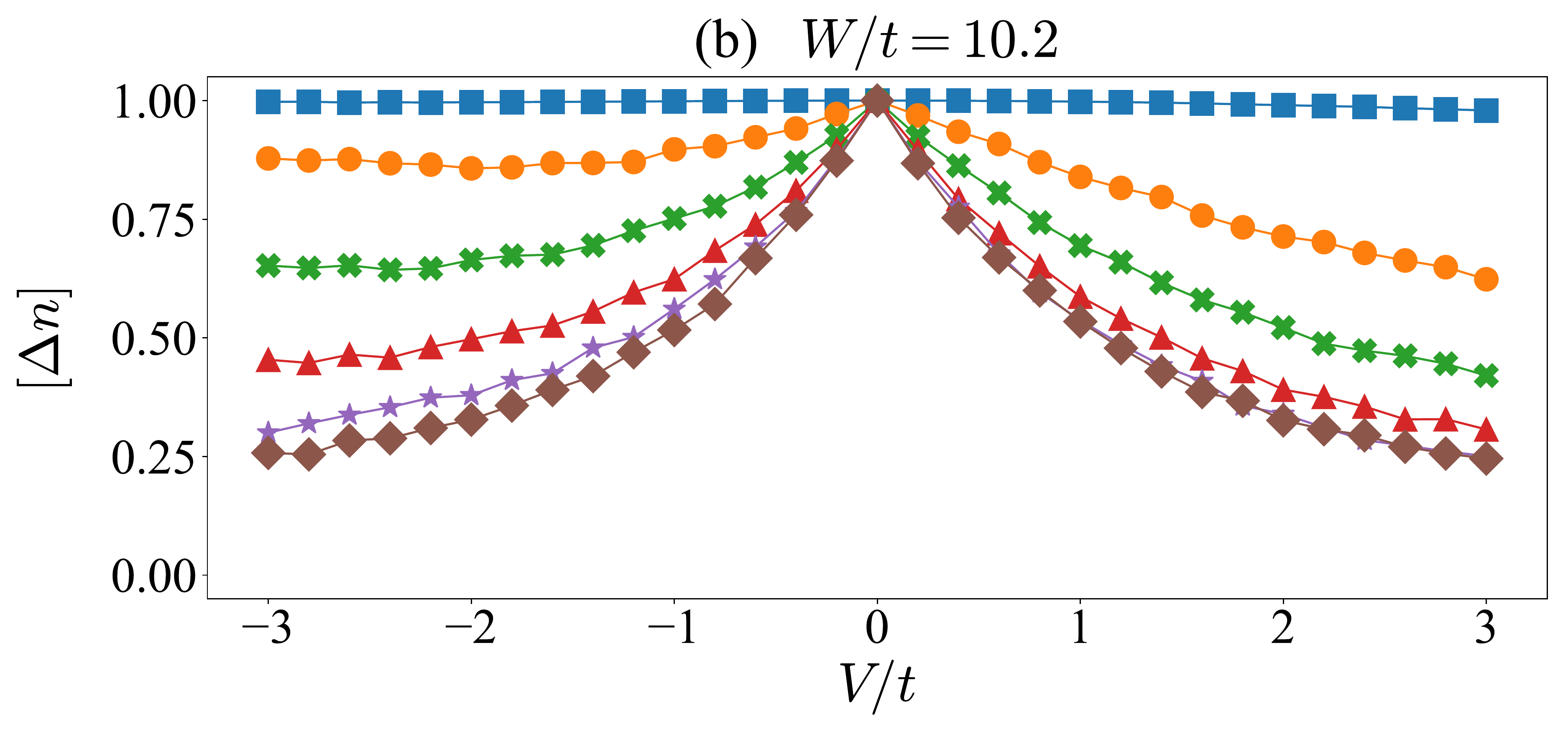}
\caption{
Cuts through the phase diagrams of Fig.~\ref{fig:deltan_W_V} at (a) $W/t=1.2$ and (b) $W/t=10.2$ for different energy densities $\epsilon=0,0.2,0.4,0.6,0.8,1$ and $L=18$.
}
\label{fig:cuts_delta_n_W_V}
\end{figure}

It is instructive to consider cuts through the $\epsilon$ versus $V$ phase diagrams at constant $W$. Examples are shown in
Figs.~\ref{fig:cuts_delta_n_W_V}(a) and (b) for $W=1.2t$ and $W=10.2t$, respectively, for several values of the energy density $\epsilon$.
Obviously, $\lbrack \Delta n\rbrack=1$ at $V=0$, independently of the value of disorder. For $\epsilon=0$, the discontinuity is
large throughout for all values of $V$. Increasing $\epsilon$ leads to a quick decay of $\lbrack \Delta n\rbrack$ away from $V=0$,
which is more significant for weak disorder, where we expect an ergodic phase everywhere except for the Anderson insulator at $V=0$
and its immediate vicinity.
Note the asymmetry with respect to $V \to -V$: at small $W/t$, the many-body states have a {\it smaller} discontinuity for $V<0$ compared to $V>0$
while the trend is opposite for strong disorder (this applies to small energy densities). 
While the asymmetry at small $W/t$ is a consequence of the ground-state phase diagram, the (inverted) asymmetry between $V$ and $-V$
at large $W/t$ can be understood based on the arguments given in \cite{Mondragon-Shem2015}. 
Their result is that ferromagnetic states should be more susceptible to localization than antiferromagnetic ones, which is reflected
in the dependence of $\lbrack \Delta n \rbrack $ on $V/t$ at small energy densities but large $W/t$ (compare Fig.~\ref{fig:deltan_W_V}(b)).

\begin{figure}[t]
\includegraphics[width=\columnwidth]{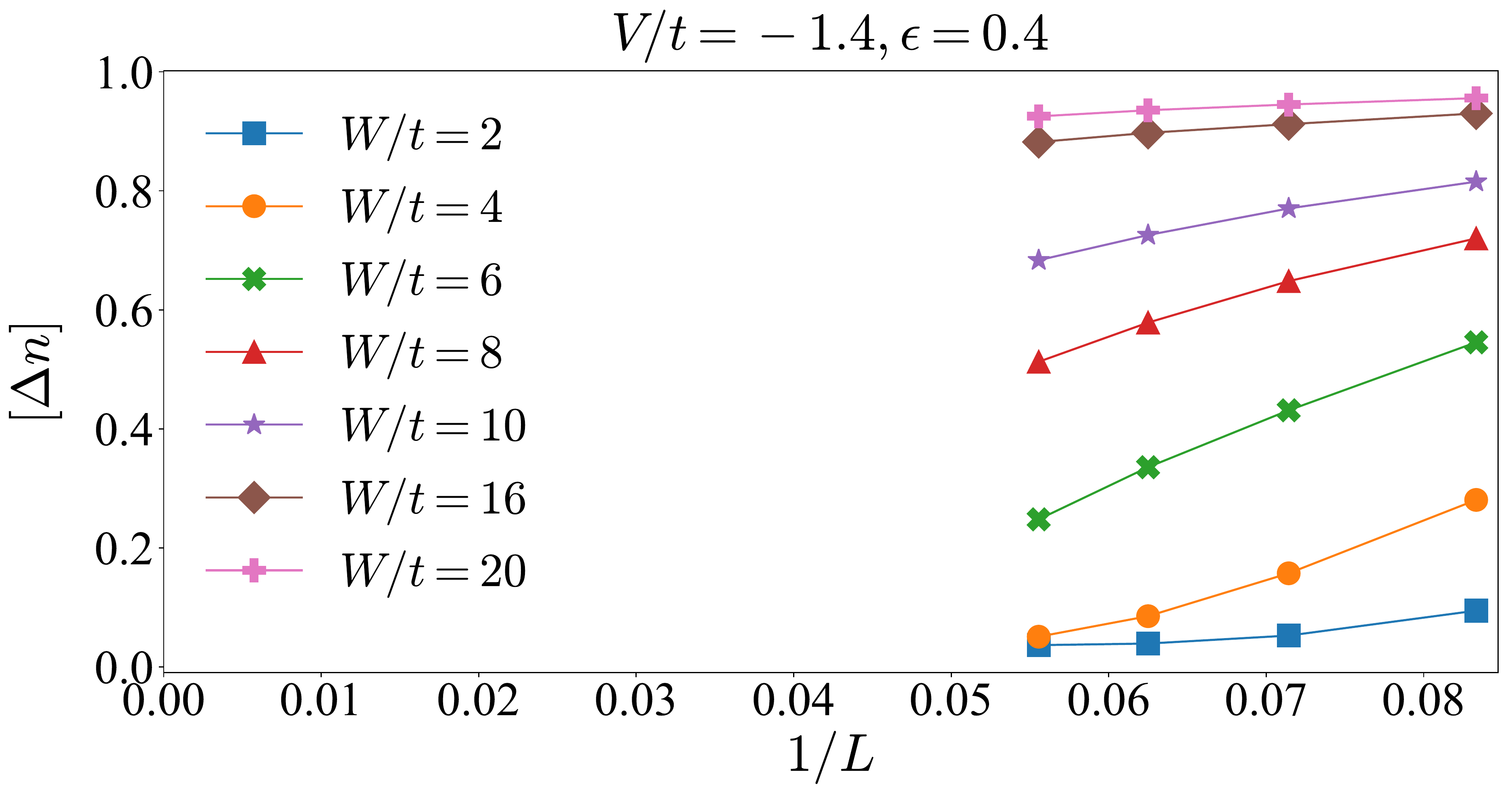}
\caption{
Finite-size dependence of the occupation-spectrum discontinuity at $V=-1.4t$ and $\epsilon=0.4$ for several disorder strengths
$W/t=2,4,6,\dots, 16,20$.
}
\label{fig:delta_n_L}
\end{figure}

Clearly, due to the small system sizes accessible to exact diagonalization, there is need to check for the robustness of the qualitative
trends as $L$ is varied. We present such an analysis in Fig.~\ref{fig:delta_n_L} for parameters in the middle of the Luttinger-liquid phase ($V=-1.4t$)
but for an elevated energy density $\epsilon=0.4$ and $2 \leq W/t \leq 20$. 
At least for the smallest values of $W/t=2,4$, the data clearly suggest that $\lbrack \Delta n \rbrack \to 0$, indicative of a phase with 
Fock-space delocalization and a continuous occupation spectrum as expected for the ergodic phase.
For $W/t \gtrsim 10$, the data seem to extrapolate to a finite value, as expected for the MBL phase \cite{Bera2015}.
These finite-size trends clearly establish  that there are two different phases separated by a Fock-space 
delocalization transition.

\subsection{Natural orbitals and inverse participation ratio}

The observation of a large occupation-spectrum discontinuity implies Fock-space localization \cite{Bera2015} but not necessarily
localization in real space. To demonstrate that in the MBL case, both features come about simultaneously, we can analyze the 
eigenstates of the OPDM as well, the natural orbitals \cite{Bera2015}.

\begin{figure}[t]
\includegraphics[width=\columnwidth]{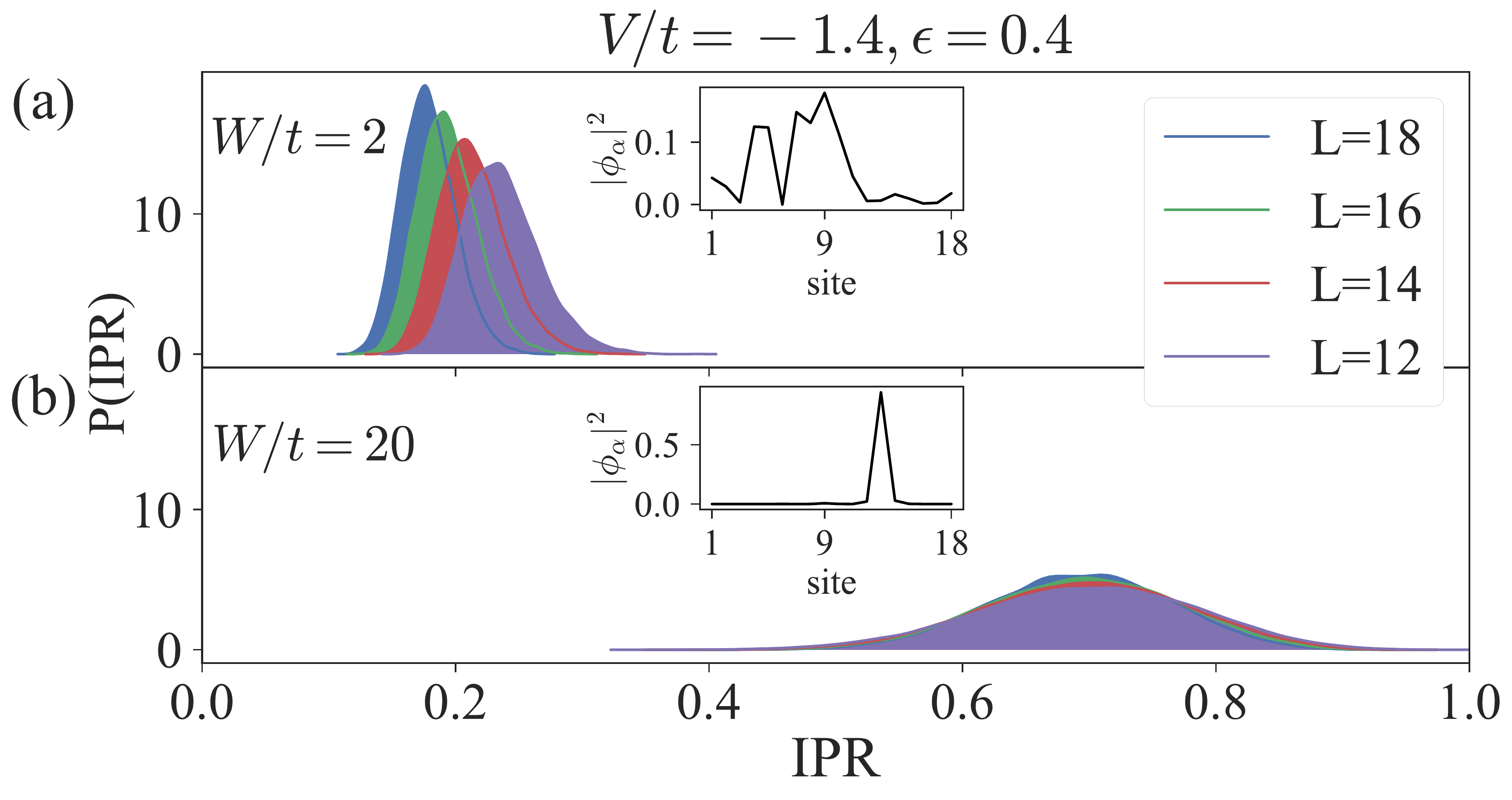}
\caption{
Distributions of inverse participation ratio in (a) ergodic phase ($W/t=2$) and (b) localized phase ($W/t=20$) for $L=18$ and $V=-1.4t$. 
We show examples of natural orbitals in both phases in the insets. The results provide evidence for real-space localization in the MBL phase. 
}
\label{fig:ipr}
\end{figure}

Figure~\ref{fig:ipr} shows the full distribution of the inverse participation ratio defined in Eq.~\eqref{eq:ipr} in its main panels for (a) the ergodic phase and (b)
the MBL phase, both at attractive interactions. Before discussing the IPR, it is illustrative to plot individual natural orbitals, which are shown in the 
insets: the one in the ergodic phase appears to be extended, while the one for the MBL phase is obviously  strongly localized in real space.

The full analysis of the distributions $P(\mbox{IPR})$ of the IPR supports this picture. In the ergodic phase, the typical value of the IPR moves to zero with $1/L$
as $L$ increases and the distributions become narrower at the same time. Conversely, in the MBL case, the distributions are broad, peaked around a large IPR value
and practically $L$-independent. This is in agreement with the results reported in \cite{Bera2015,Bera2017} for repulsive interactions and immediately
visualizes the localization of the quasi-particles (or l-bits) in real space by approximating their single-particle content via diagonalizing the OPDM \cite{Bera2017}.

\subsection{Von-Neumann entropy}

\begin{figure}[t]
\includegraphics[width=\columnwidth]{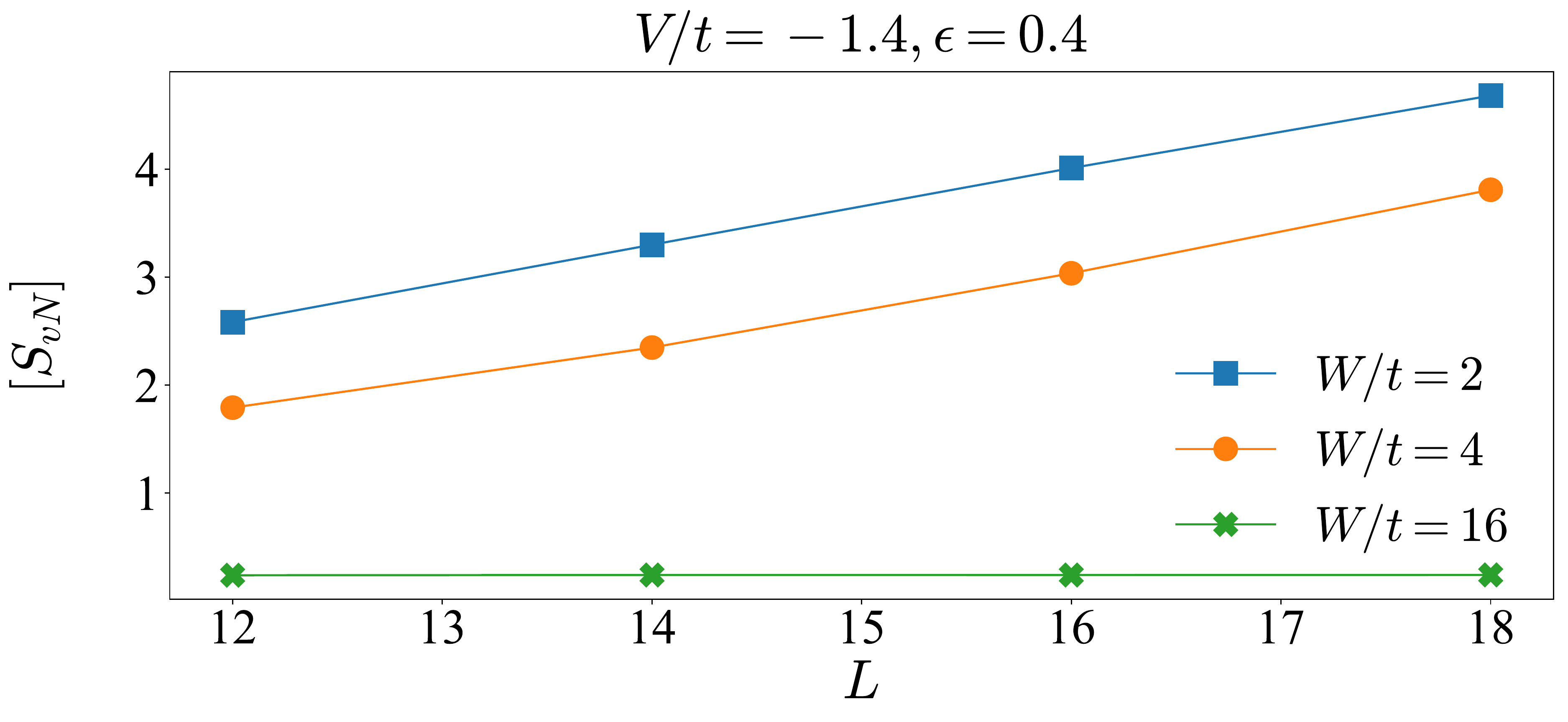}
\caption{
Half-cut von-Neumann entropy versus system size in the ergodic $(W/t=2,4)$ and MBL $(W/t=16)$ phase at $V/t=-1.4$ and $\epsilon=0.4$.
}
\label{fig:svn_half}
\end{figure}

As an alternative measure of the transition, we analyze the von-Neumann entropy, which we expect to exhibit a volume law in the 
ergodic phase but  an area-law in the MBL phase. We verify this expectation by plotting the half-cut entanglement entropy versus system size
in Fig.~\ref{fig:svn_half} for two points deep in the ergodic and deep in the MBL region. 

\begin{figure}[t]
\includegraphics[width=\columnwidth]{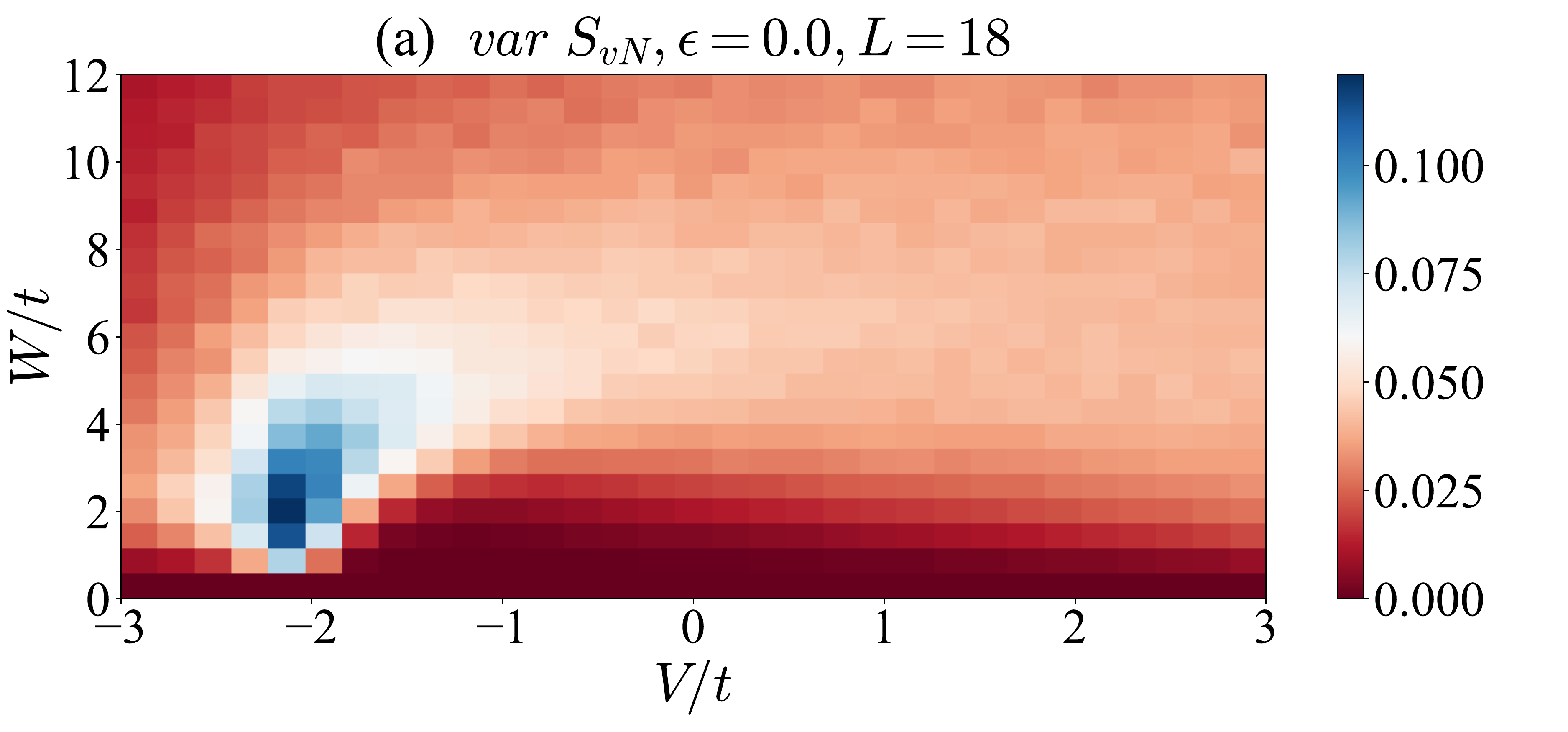}
\includegraphics[width=\columnwidth]{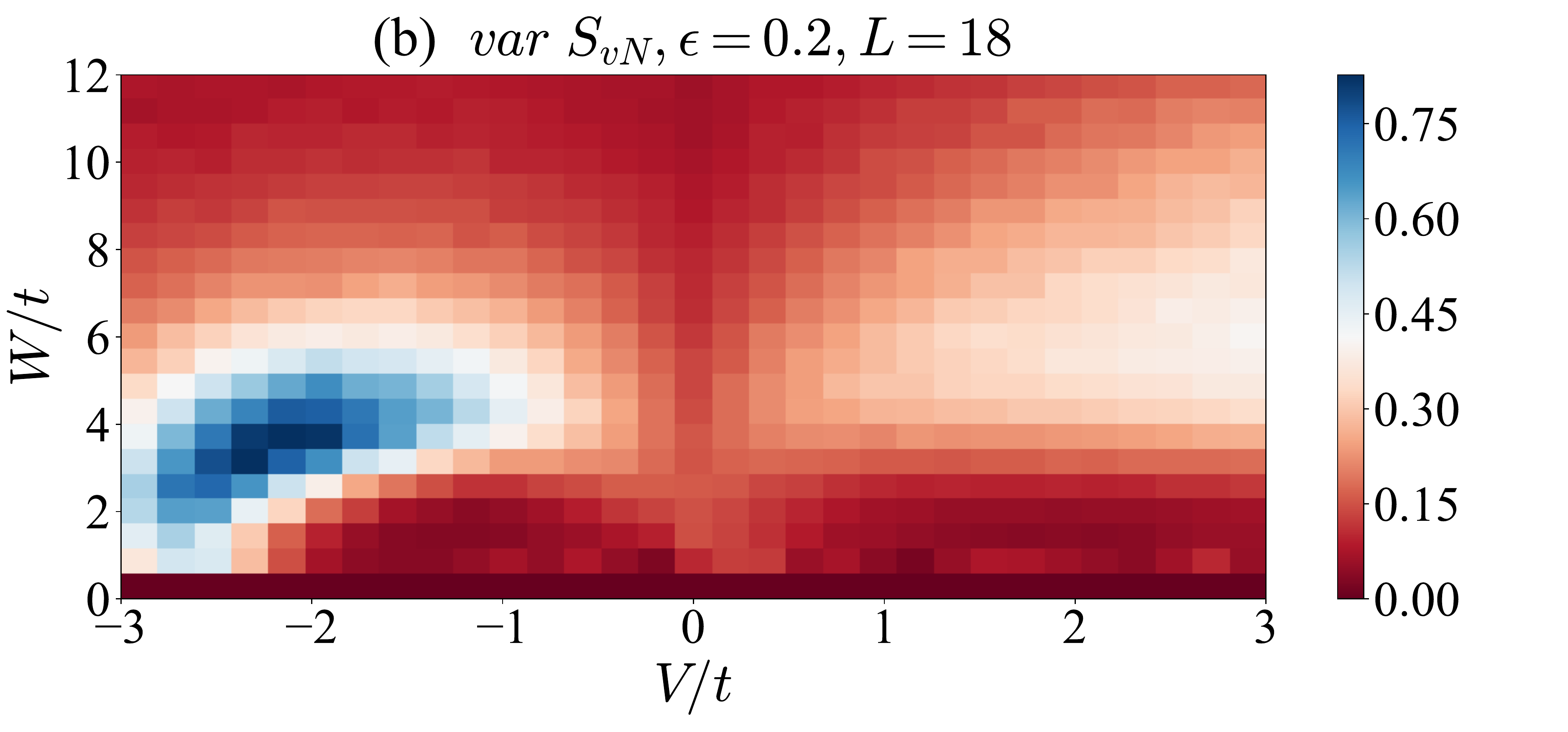}
\includegraphics[width=\columnwidth]{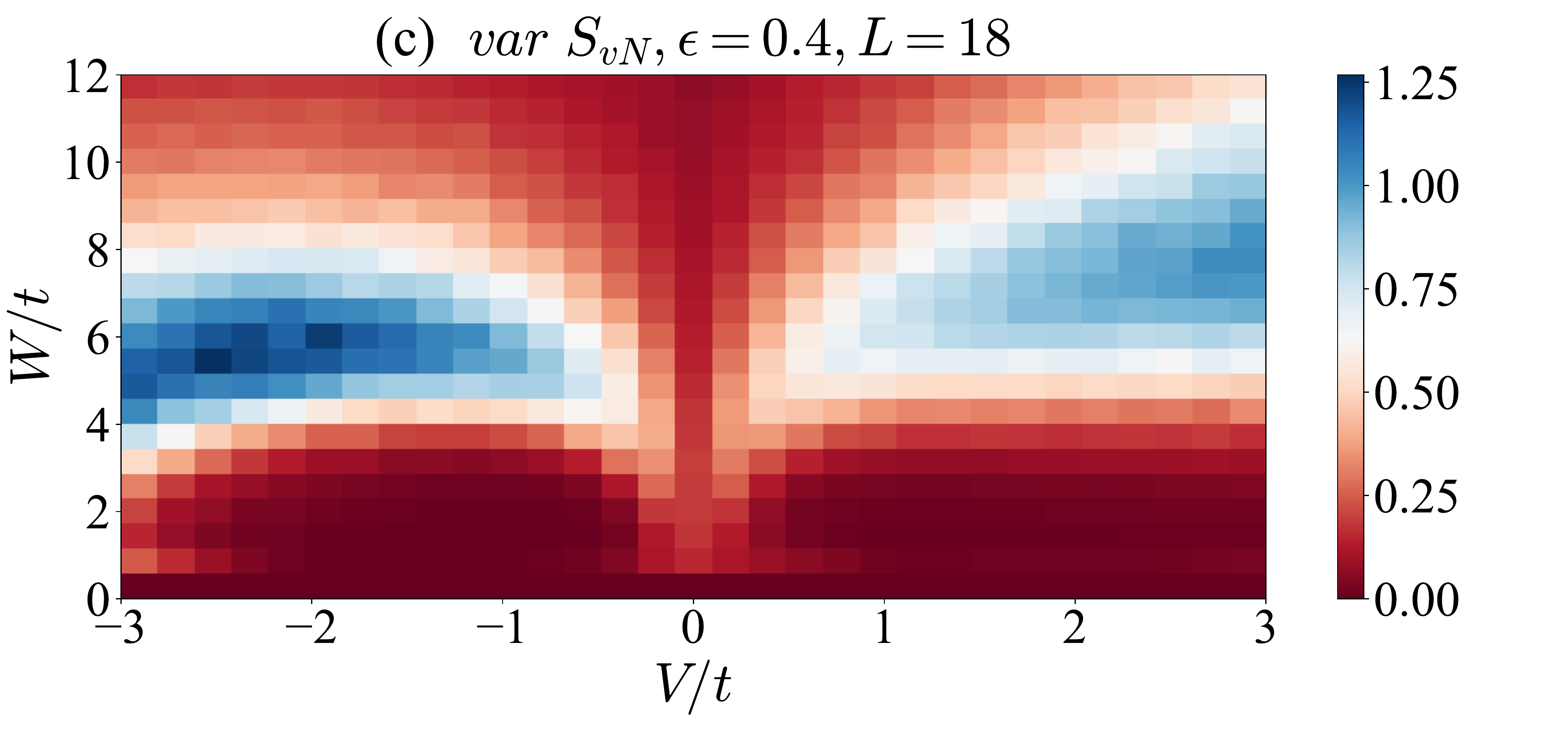}
\includegraphics[width=\columnwidth]{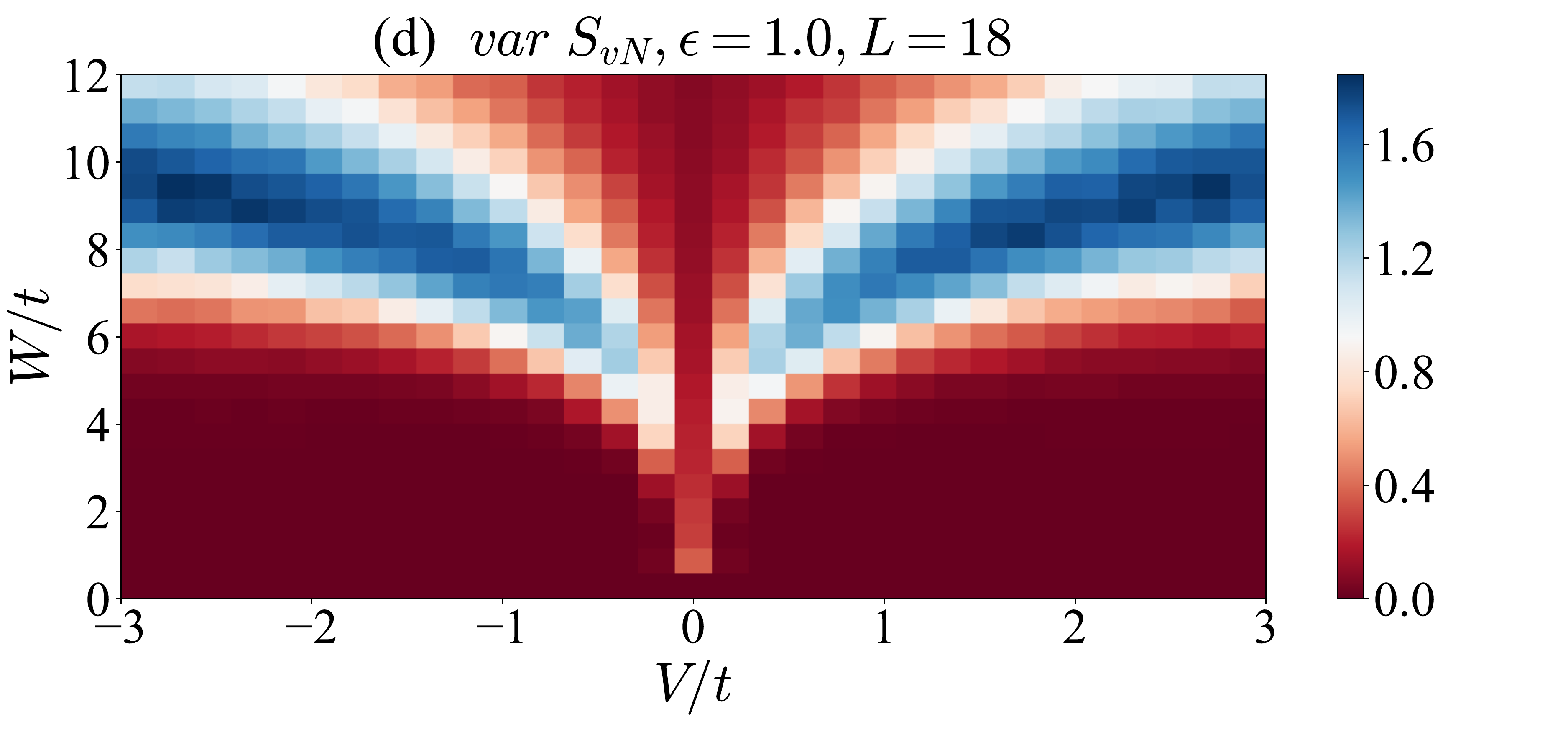}
\caption{Variance of the von-Neumann entropy in the $W$ versus $V$ plane for (a) $\epsilon=0 $, (b) $\epsilon=0.2$, (c) $\epsilon=0.4 $, (d) $\epsilon=1 $, $L=18$.
}
\label{fig:svn_W_V}
\end{figure}

The location of the transition can be estimated from the position of the maximum of sample-to-sample fluctuations
of the von-Neumann entropy, $\mbox{var}\, S_{\rm vN}$ \cite{Kjaell2014}. We plot this quantity in Fig.~\ref{fig:svn_W_V}
for the same parameters as in Figs.~\ref{fig:deltan_W_V}(a)-(d).
The behavior of $\mbox{var}\, S_{\rm vN}$  confirms the picture obtained from the occupation-spectrum discontinuity, at least
for $\epsilon>0$. As energy density increases, the transition between the weak-disorder delocalized and the strong-disorder localized phase moves to larger values of 
$W/t$. In the ground state, i.e., at $\epsilon=0$, and for small system sizes, only the phase boundary  at large negative values of $V$ is resolved by a 
clear maximum in $\mbox{var}\, S_{\rm vN}$, while this is not the case for the transition at small negative values.
We believe this to be a consequence of (i) the large single-particle localization length on the attractive side and (ii) the different 
nature of the Luttinger-liquid to localization transition at $\epsilon=0$ (see the discussion in \cite{Ristivojevic2012,Doggen2017}).

By comparison of Fig.~\ref{fig:svn_W_V}(d) and the corresponding data for the occupation-spectrum discontinuity $\lbrack \Delta n\rbrack $
shown in Fig.~\ref{fig:deltan_W_V}(d), one realizes that {\it at a fixed system size} var$\,S_{\rm vN}$ peaks at significantly lower values than
where $\lbrack \Delta n\rbrack $ exhibits a drop to (approximately) zero. First, this simply reflects that the actual transition point can only
be estimated from a finite-size scaling analysis (see the discussion in \cite{Khemani2017} though). Second, different quantities 
exhibit different finite-size deviations from large-$L$ behavior. The fact that var$\,S_{\rm vN}$ has a stronger finite-size dependence
with the maximum sitting well below the actual transition point is consistent with the observation of other studies (see, e.g., \cite{Khemani2016}).

\subsection{Level-spacing distribution}

\begin{figure}[!t]

\includegraphics[width=\columnwidth]{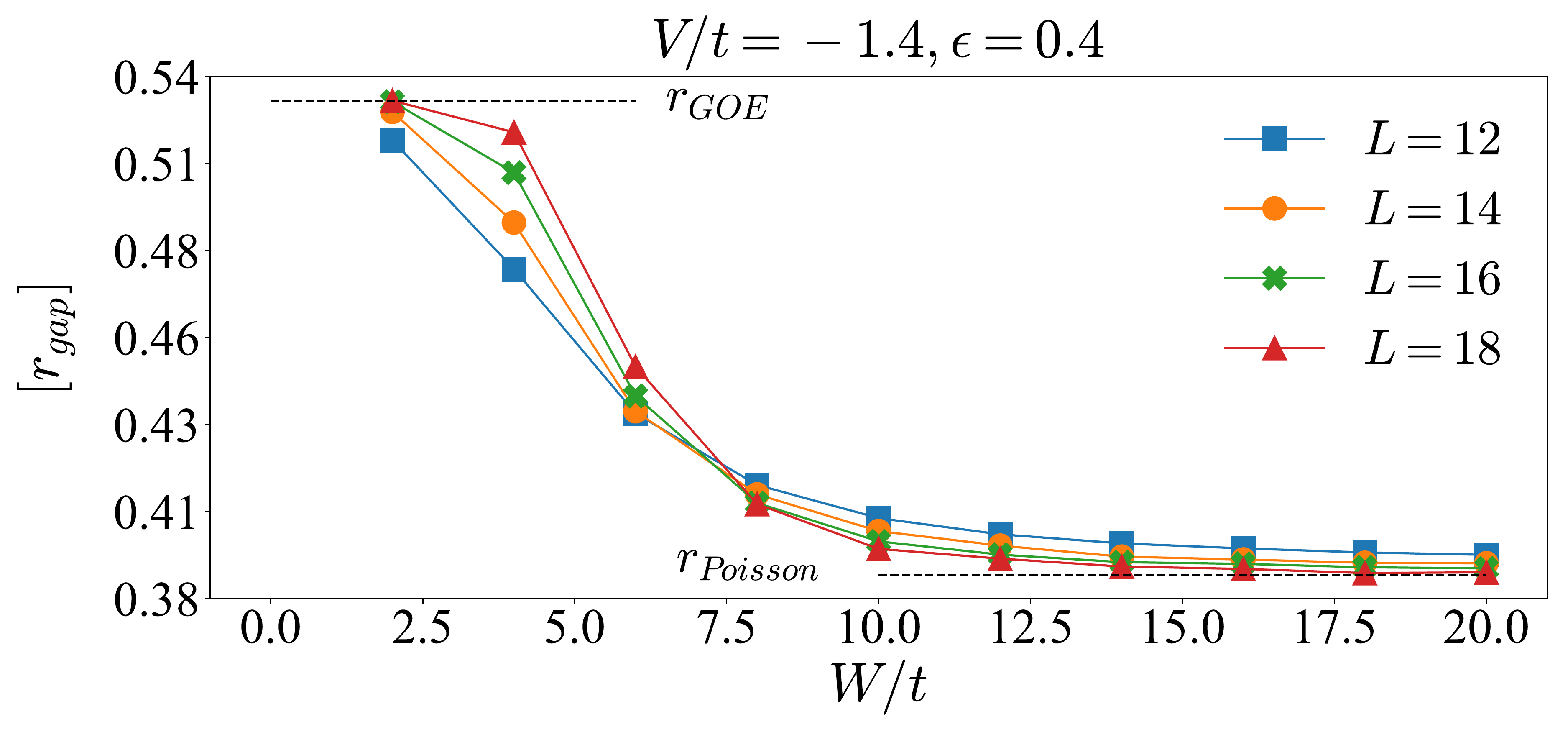}
\caption{
Adjacent gap ratio as a function of disorder strength at energy density $\epsilon = 0.4$ and interaction strength $V = -1.4t$. 
The result matches the theoretically predicted adjacent gap ratio $r_{\text{GOE}} = 0.5307$ and $r_{\text{Poisson}} = 0.3863$. Error bars are smaller than the symbol size.
}
\label{fig:gap_ratio}
\end{figure}

An example for the $W$-dependence of the adjacent-gap ratio $r_{\rm gap}$ is presented in Fig.~\ref{fig:gap_ratio}
for $V=-1.4t$ and $\epsilon=0.4$.
As expected \cite{Oganesyan2007}, $r_{\rm gap}$ approaches the values expected for a Wigner-Dyson distribution of the underlying level spacings (i.e., $r_{\rm gap}=0.3863$) at small values of $W/t$
and goes to $r_{\text{gap}} = 0.3863$ as $W$ increases, which takes the system into the MBL phase.
The data suggests that the transition is at $6 \lesssim W/t\lesssim 8$ for the parameters of the figure. 

\subsection{Energy-density versus interaction-strength phase diagram}
\begin{figure}[!t]
\includegraphics[width=\columnwidth]{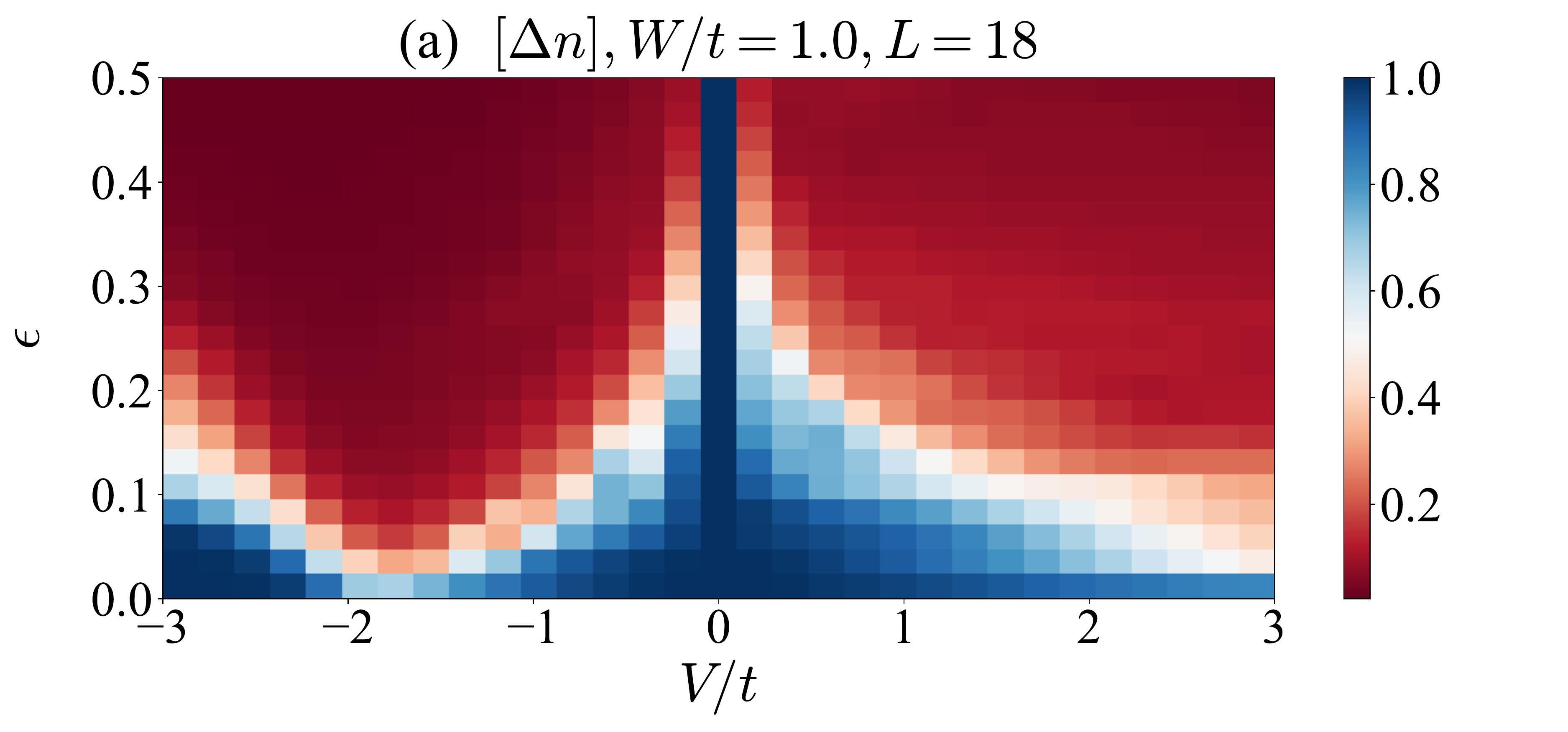}
\includegraphics[width=\columnwidth]{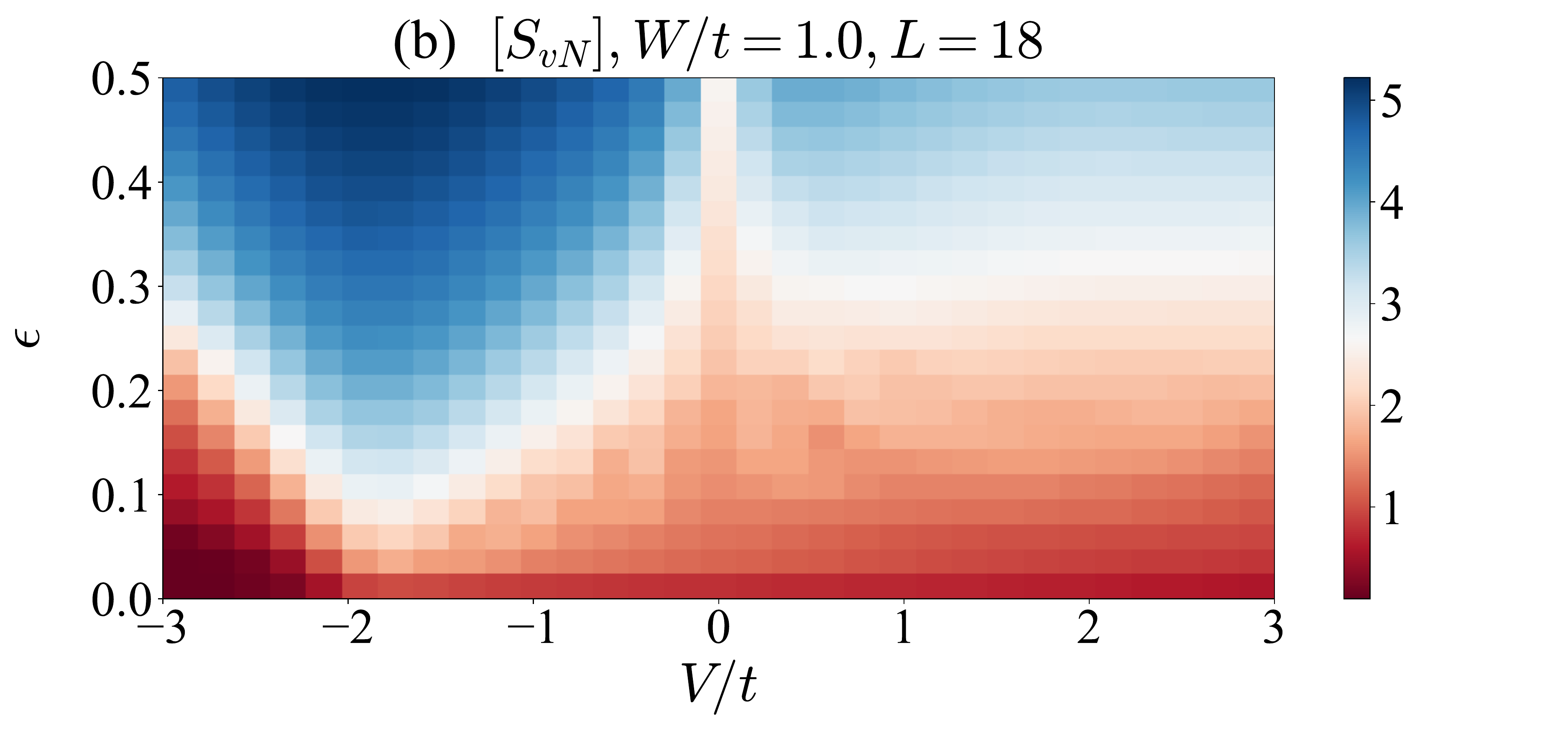}
\includegraphics[width=\columnwidth]{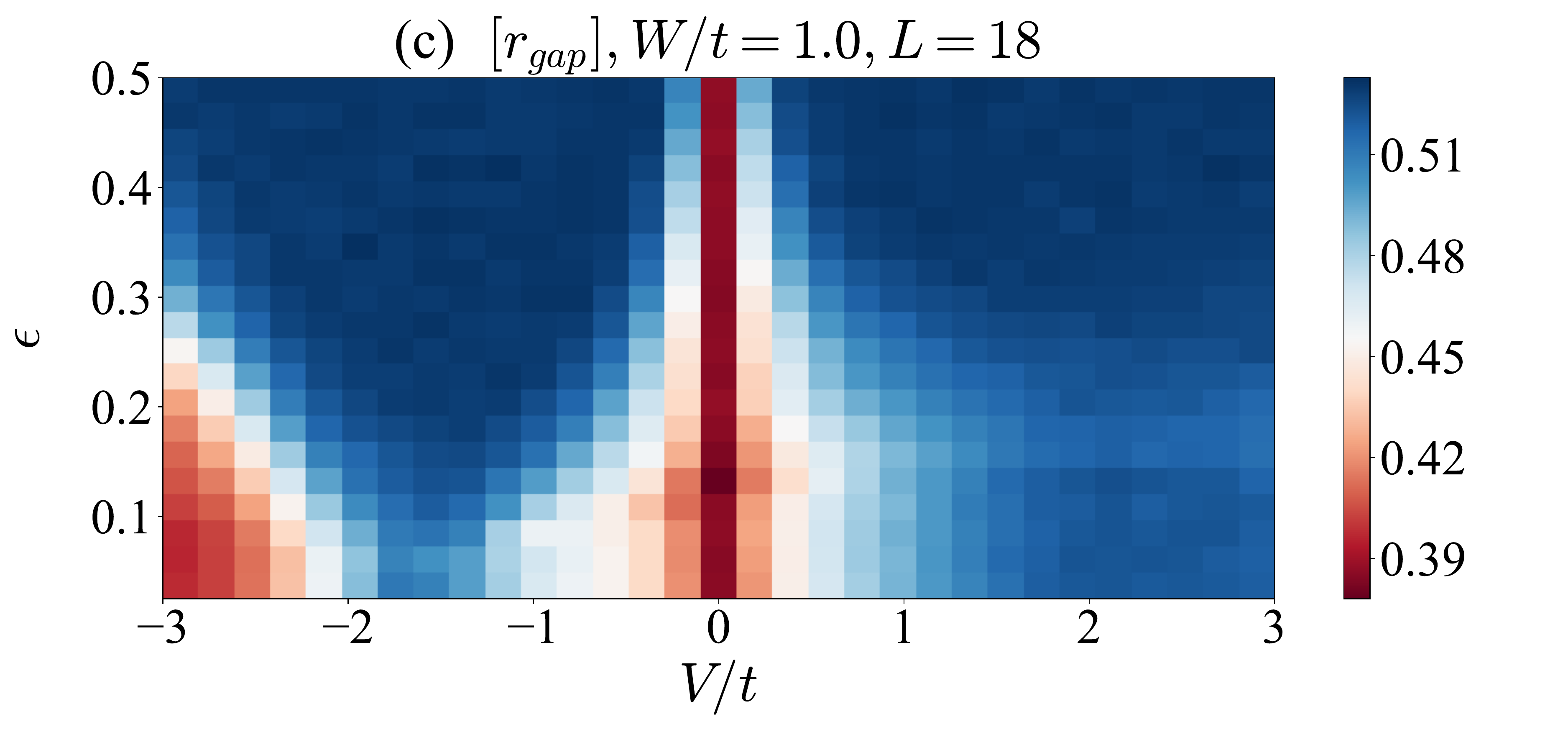}
\caption{Energy-density versus interaction strength phase diagram at weak disorder: (a) OPDM occupation discontinuity, (b) half-chain cut von Neumann entropy and 
(c) adjacent gap ratio. 
The  delocalized phase expands as the energy density increases. The predictions from $\lbrack {\Delta n}\rbrack$, $\lbrack S_{VN} \rbrack$, and $\lbrack r_{gap} \rbrack$ are qualitatively consistent. The delocalized phase is not that obvious in the ground-state region for this system size, due to the comparably much smaller discontinuities at finite energy densities. Note that (c) does not include the ground-state region and hence
does not show the mobility edge on the $V>0$ side. 
}
\label{fig:phase_diag_e_V}
\end{figure}

The main result of our work is an energy-density versus interaction-strength phase diagram 
at weak disorder. We present such  diagrams for $W=t$ in Figs.~\ref{fig:phase_diag_e_V}(a)-(c),
derived from the occupation-spectrum discontinuity, the half-cut von-Neumann entropy, and the adjacent-gap ratio $r_{\rm gap}$, respectively.
All three quantities give a qualitatively consistent picture: at negative $V<0$, the ergodic phase extends down to $\epsilon=0$ and thus adiabatically
connects to the Luttinger-liquid phase, while for repulsive interactions, there is always a mobility edge as $\epsilon$ increases.
Moreover, we can thus rule out the presence of an inverted mobility edge from the data for $L\leq 18$ sites in our model.
The phase diagram further illustrates the asymmetry under changing the sign of $V$.

We here present results for a fixed system size $L=18$, while in principle, one can carry out a finite-size analysis
and even attempt a finite-size scaling collapse of the data \cite{Luitz2015,Kjaell2014}. Such analyses were carried out
for $V>0$ but the resulting exponents are inconsistent with rigorous bounds \cite{Harris1974,Chayes1986,Chandran2015}.
Hence, we here do not pursue this strategy since our system sizes are not larger than what was used in \cite{Luitz2015,Kjaell2014}.
Moreover, the question of what the universality class of the delocalization-localization transition  in models such as ours is, is still
a topic of ongoing research \cite{Kjaell2014,Vosk2015,Potter2015,Devakul2015,Zhang2016,Khemani2016,Khemani2017}

By inspection of Figs.~\ref{fig:phase_diag_e_V}(a)-(c) one realizes that the occupation-spectrum discontinuity and the adjacent-gap ratio
provide the best resolution of the phase diagram as both quantities have small finite-size effects deep in the respective phase. Note that
we plot the adjacent-gap ratio only down to $\epsilon \geq 0.025t$.
The statistics of the adjacent gap ratio does not obey the predictions \cite{Oganesyan2007,Pal2010} as $\epsilon \rightarrow 0$, where finite-size effects are expected to be the largest.

\section{Conclusion}
\label{sec:sum}

In this work we studied the finite energy-density phase diagram of spinless fermions in a one-dimensional lattice
with attractive interactions and uncorrelated disorder. Our numerical results illustrate an  asymmetry between
the $V<0$ and $V>0$ phase diagrams, rooted in the existence of a delocalized phase in the ground state of the system with 
attractive interactions and weak disorder \cite{giamarchi,Schmitteckert1998}. This Luttinger-liquid phase connects to a finite-temperature ergodic phase without
any intermediate phase transition, as the analysis of the entanglement entropy, the level-spacing distribution and the one-particle
density matrix occupation spectrum indicates. This conclusion rules out the existence of an inverted mobility edge in our model, which may still
exist in other disordered systems with superfluids present in the ground state, such as the Bose-Hubbard model. 
In fact a series of very recent studies  \cite{Singh2017,Sierant2017,Sierant2017a} indicates that there is an inverted mobility edge in the one-dimensional Bose-Hubbard model.
This different behavior compared to our case of fermions with attractive interactions could be traced back to 
two observations. First, in the case of bosons, disorder makes the stability region of the superfluid (as a function of $W$ and interaction strength) larger  than in the clean case. Second, the Bose-Hubbard model is prone to a sort of dynamical localization even in the absence of disorder:
composite opbjects of multiple bosons themselves can get localized in individual sites and may have very large decay times
in the limit of strong interactions \cite{Grover2014,Schiulaz2014,Deroeck2014}. Increasing energy density has been suggested as one mechanism to induce such heavy
objects, which in the presence of disorder would get localized. 

We also revisited the ground-state case, with two main observations. First, the Luttinger-liquid region features a significantly smaller
OPDM occupation-spectrum discontinuity compared to the localized ground-state phases on finite systems, as expected for a Luttinger-liquid,
which
has a vanishing discontinuity in the thermodynamic limit. The  localized regions, on the other hand, 
have a Fock-space localized ground-state wavefunction, just as in the finite-energy density case \cite{Bera2015,Bera2017}.
Second, we used functional renormalization group simulations to compute the conductance at zero temperature, which also resolves
the delocalized ground-state region. While the FRG simulations for the conductance are computationally cheap and can be pushed to
system sizes as large as $L\sim 10^6$, the phase boundaries, in particular, at large negative interaction strengths $V \lesssim -2t$, are quantitatively
different from the results of other methods (see, e.g., \cite{Doggen2017}). Thus, improvements in the FRG scheme are necessary to render the method accurate
at larger interactions strengths as well, which poses an interesting direction for further method development.

{\it Acknowledgment} We thank D. Luitz for useful discussions and comments on a previous version of the manuscript.
The  work of F.H.-M. was performed in part at Aspen Center for Physics, which is supported by National Science Foundation grant PHY-1607611.'
The hospitality of the  Aspen Center for Physics is gratefully acknowledged.
 F.D. and F.H.-M. were supported by the Deutsche Forschungsgemeinschaft (DFG) via Research Unit FOR 1807 under grants No. HE 5242/3-1 and  No. HE 5242/3-2.   

\bibliography{references}
\end{document}